\documentclass[apj]{emulateapj}
\slugcomment{{\sc Accepted to ApJS:} August 8, 2006}
\def\gtorder{\mathrel{\raise.3ex\hbox{$>$}\mkern-14mu
                \lower0.6ex\hbox{$\sim$}}}
\def\ltorder{\mathrel{\raise.3ex\hbox{$<$}\mkern-14mu
                \lower0.6ex\hbox{$\sim$}}}

\shorttitle{Time-Dependent Radiative Cooling}
\shortauthors{Gnat \& Sternberg}

\begin{document}
\title{Time-Dependent Ionization in Radiatively Cooling Gas}
\vspace{1cm}
\author{Orly Gnat$^{1,2}$ and Amiel Sternberg$^1$}
\vspace{0.5cm}
\affil{$^1$~School of Physics and Astronomy and the Wise Observatory,
        The Beverly and Raymond Sackler Faculty of Exact Sciences,
        Tel Aviv University, Tel Aviv 69978, Israel \\
       $^2$~Theoretical Astrophysics, California Institute of Technology, 
        MC 130-33, Pasadena, CA 91125.}
\email{orlyg@tapir.caltech.edu}

\begin{abstract}
We present new computations of the equilibrium and non-equilibrium 
cooling efficiencies and ionization states for
low-density radiatively cooling gas containing cosmic abundances of the
elements H, He, C, N, O, Ne, Mg, Si, S, and Fe. We present results for 
gas temperatures between $10^4$ and $10^8$~K, 
assuming dust-free and optically thin conditions, and no external radiation.
For non-equilibrium cooling we solve the coupled time-dependent
ionization and energy loss equations for a radiating
gas cooling from an initially hot, $\gtrsim 5\times 10^6$~K,
equilibrium state, down to $10^4$ K.
We present results for
heavy element compositions ranging from $10^{-3}$ to
$2$ times the elemental abundances in the Sun.
We consider gas cooling at
constant density (isochoric) and at constant pressure (isobaric). 
We calculate the critical column densities and
temperatures at which radiatively cooling clouds make
the dynamical transition from isobaric to isochoric evolution.
We construct ion ratio diagnostics for
the temperature and metallicity in radiatively cooling gas.
We provide numerical estimates for the maximal cloud column
densities for which the gas remains optically thin to
the cooling radiation.
We present our computational results in convenient on-line
figures and tables (see http://wise-obs.tau.ac.il/$\sim$orlyg/cooling/).
\end{abstract}


\section{Introduction}
\label{introduction}
The collisionally controlled ionization states and radiative
cooling rates of hot ($10^4-10^8$~K) low-density ``coronal'' gas clouds
are crucial quantities in the study of the diffuse interstellar 
and intergalactic medium, including hot gas in supernova remnants, 
the Galactic halo, active galactic nuclei, galaxy groups and clusters, 
and warm/hot ``baryonic reservoirs'' in intergalactic clouds and filaments. 
In this paper we present new computations of the equilibrium and
non-equilibrium cooling efficiencies and ionization states 
for radiatively cooling gas clouds with
temperatures between $10^4$ and $10^8$~K. Our main emphasis is on
the non-equilibrium cooling and ionization states 
that occur when an initially hot gas cools radiatively below
$\sim 5\times 10^6$ K.
 Below this temperature, cooling can become
rapid compared to ion-electron recombination, and the gas at any
temperature tends to remain ``over-ionized'' compared to
gas in collisional ionization equilibrium (CIE).
Because the energy losses below $\sim 10^7$~K are dominated by
atomic and ionic line emissions from many species, the non-equilibrium
cooling rates also differ
(are suppressed for over-ionized gas) compared to
equilibrium cooling. The ``recombination lag'' and
departures from equilibrium depend
on the abundances of the heavy elements, and we present
results for compositions ranging from $10^{-3}$ to 2 times
the elemental abundances in the Sun.
We include H, He, C, N, O, Ne, Mg, Si, S, and Fe,
and we calculate the ion fractions as functions of temperature,
for both equilibrium and non-equilibrium cooling,
for all ionization stages of these elements.

The non-equilibrium radiative cooling of hot ($>10^4$ K),
highly-ionized gas, with emphasis on the
associated time-dependent ionization of the metal ions, is a classical
problem first investigated by Kafatos (1973),
and subsequently by several authors including Shapiro \& Moore (1976),
Edgar \& Chevalier (1986), Schmutzler \& Tscharnuter (1993),
Sutherland \& Dopita (1993), and Smith et al.~(1996).
Time dependent cooling in colder ($T\lesssim 10^4$~K) 
neutral hydrogen gas was studied earlier by Bottcher et al.~(1970),
Jura \& Dalgarno (1972), and Schwarz et al.~(1972) (see also
Dalgarno \& McCray [1972]).
These investigations demonstrated that below a few $10^6$~K
recombination lags can indeed lead to significant departures from
CIE states and the associated equilibrium cooling rates. Computations 
of hot gas cooling efficiencies and ionization states 
assuming CIE have been presented in many papers
dating back to House (1964), Tucker \& Gould (1966),
Allen \& Dupree (1969), Cox \& Tucker (1969),
Jordan (1969), followed by Raymond et al.~(1976),
Shull \& van Steenberg (1982), Gaetz \& Salpeter (1983),
Arnaud \& Rothenflug (1985), Boehringer \& Hensler (1989),
and more recently Sutherland \& Dopita (1993),
Landi \& Landini (1999), and Benjamin et al.~(2001).

In this paper we reexamine the fundamental problem of non-equilibrium ionization
in a time-dependent radiatively cooling gas. 
We focus on the behavior for pure radiative cooling
with no external sources of
heat or photoionization. 
Recombination lags and non-equilibrium ionization also occur
in an initially hot gas undergoing rapid expansion and adiabatic cooling 
(Breitschwerdt \& Schmutzler 1994; 1999). We do not consider
such dynamical effects here.

Our work is motivated by ultraviolet (UV) and X-ray
metal absorption-line spectroscopy of hot gas
in the Galactic halo (e.g., Sembach \& Savage 1992; Spitzer 1996),
including, more recently, in discrete ionized ``high-velocity clouds''
(Savage et al.~2002; Fox et al.~2004, 2005, Fox, Savage, \& Wakker~2006, see also
Sternberg, McKee, \& Wolfire~2002; Gnat \& Sternberg 2004;
Maller \& Bullock 2004). We are also motivated by
recent {\it Hubble Space Telescope (HST)},
{\it Far Ultraviolet Spectroscopic Explorer (FUSE)},
{\it Chandra}, 
and {\it XMM-Newton}, detections and observations of hot gas
around the Galaxy (Collins et al.~2005; Fang et al.~2006), and/or of gas
that may be part
of a 10$^5$-10$^7$ K ``warm/hot intergalactic medium'' (WHIM)
(Tripp, Savage, \& Jenkins~2000; Shull, Tumlinson, \& Giroux~2003; 
Richter et al.~2004; Sembach et 
al.~2004; 
Soltan, Freyberg, \& Hasinger 2005;
Nicastro et al.~2005; Savage et al.~2005, Williams et al.~2006;
see however Rasmussen et al.~2006).
This includes the temperature range where departures from
equilibrium cooling are expected to be important.
An intergalactic shock heated WHIM is expected to be a major reservoir
of baryons in the low redshift universe (Cen \& Ostriker 1999; Dav{\'e} et
al.~2001; Furlanetto et al.~2005).
Important ions for the detection of
such gas include \ion{C}{4}, \ion{N}{5},
\ion{O}{6}, \ion{O}{7}, \ion{O}{8}, \ion{Si}{3}, \ion{Si}{4},
\ion{Ne}{8}, and \ion{Ne}{9}.

The early discussions of Kafatos (1973) and Shapiro \& Moore (1976)
focused on time-dependent isochoric (constant density)
cooling at a single metallicity.
Sutherland \& Dopita (1993) calculated isobaric 
(constant pressure) cooling
efficiencies for a range of metallicities, but presented limited
results for the time-dependent ion-fractions, and
did not consider isochoric cooling.
Schmutzler \& Tscharnuter (1993) 
considered isochoric cooling for a range of 
metallicities, and presented ion-fractions only for solar metallicity gas.
They presented limited results for isobaric cooling.
In our calculations,
we use up-to-date atomic data, and we present complete results
for the time-dependent ion fractions and cooling efficiencies, 
for both isochoric and isobaric evolutions,
for a wide range of metallicities.
Because of the ``$PdV$'' work, isobaric cooling
is less rapid than isochoric cooling, so 
departures from equilibrium are somewhat smaller in the isobaric case.

We consider metallicities ranging from $10^{-3}$ to twice the solar abundances
of the heavy elements.
In our calculations we assume that any initial dust is
rapidly destroyed (e.g.~by thermal sputtering) on a
time-scale that is short compared to the initial cooling rate
(Smith et al.~1996), and we consider dust-free cooling at
fixed gas-phase elemental abundances. 
The sensitivity of the cooling efficiencies to the metallicity
was studied by Boehringer \& Hensler (1989) for equilibrium cooling,
and by Schmutzler \& Tscharnuter (1993) and Sutherland and Dopita (1993)
for non-equilibrium conditions as well.
Time-dependent effects are enhanced (diminished)
as the cooling efficiencies are increased (reduced) for larger
(smaller) gas-phase abundances of the heavy elements. 
Thus, recombination lags
are more significant as the metallicity is increased.
For a low-density gas, the non-equilibrium ion fractions and cooling
efficiencies are independent of the gas 
density or pressure, because the ratio of the cooling and recombination times
is independent of density. The density scales out of the problem.
If the gas is assumed to cool, at either constant density
or constant pressure, from an
initial equilibrium state at high temperature,
the only remaining free
parameter in the problem is the metallicity $Z$. 

We present results for the optically thin limit in which 
reabsorption of diffuse ``self-radiation'' may be neglected.
We verify previous findings that trapping of hydrogen and helium
recombination radiation, 
in optically thick clouds has only a small effect on the
cooling and resulting ionization states.
We also provide estimates
of the maximal cloud column densities for which reabsorption of
radiation from the metals (lines and continua) may be neglected
in computations of the gas cooling efficiencies.
Elsewhere, we will consider the effects of
external fields
on the non-equilibrium evolution of radiatively cooling gas.
The inclusion of such radiation will introduce a pressure/density
dependence via the ratio of the photoionization
and recombination times.

In a separate paper we will present computations of the
integrated metal ion ``cooling columns'' produced in 
steady flows of cooling gas (e.g.~Edgar \& Chevalier 1986;
Dopita \& Sutherland 1996; Heckman et al.~2002)
including the effects of ``upstream'' cooling radiation 
on the ``downstream'' ionization states. Such flows
occur, for example, in the radiative (isobaric) cooling
layers behind steady shock waves (Shull \& McKee 1979; Draine \& Mckee 1993).

The outline of our paper is as follows.
In \S 2 we write down the basic equations that we solve in our
computations, and we describe our numerical method.
In \S 3 we discuss isochoric versus isobaric cooling,
and we derive critical cloud column densities
and temperatures at which the dynamical transition from
isobaric to isochoric cooling occurs.
In \S 4 we present our computations of the ionization states
as functions of gas temperature, for both CIE and non-equilibrium 
cooling. For time-dependent ionization we present results for
both isochoric and isobaric evolutions.
In \S 5 we present our computations of the radiative cooling efficiencies.
We compare the CIE and non-equilibrium cooling efficiencies, and
determine the temperatures below which non-equilibrium effects
become important for each assumed metallicity. 
In this paper we have generated a large number of figures and tables
containing results for metallicities ranging from $10^{-3}$ to
2 times solar.  Some of the figures and tables
appear in the main text, and some are available as online data files.
Our data sets can be used to construct ion density-ratio diagnostics
for radiatively cooling gas for a wide range of conditions.
As an example, in \S 6 we discuss the evolution of the density ratios
\ion{C}{4}/\ion{O}{6} versus \ion{N}{5}/\ion{O}{6} in
radiatively cooling gas. We summarize in \S 7.

\section{Basic Equations and Processes}
\label{physics}
We are interested in studying the evolving ionization states
of radiatively cooling gas, which is 
initially heated to some high temperature $T_0 \gtrsim 5\times 10^6$ K,
and then cools from an initial (equilibrium) ionization state.
In the absence of continued heating and ionization
the ions recombine as the gas cools, 
and the overall ionization state decreases with time.
If the gas cools faster than it recombines,
non-equilibrium effects become significant, and the gas remains over-ionized
throughout the cooling. Non-equilibrium ionization leads
to a suppression of the cooling rates. 
We compute this coupled time-dependent evolution
for clouds cooling at constant density or
constant pressure, for a wide range of metallicities.
Non-equilibrium effects are most important at high metallicity,
where the cooling times are shortest.

\subsection{Ionization}
\label{physics-ion}
In our numerical computations we follow the time-dependent 
ionization of an optically thin ``parcel'' of cooling gas, given an initial 
ionization state at an initial temperature $T_0$.
We consider all ionization stages of the elements 
H, He, C, N, O, Ne, Mg, Si, S, and Fe. The temperature-dependent
ionization and recombination processes that we include are
collisional ionization by thermal electrons (Voronov 1997),
radiative recombination (Aldrovandi \& Pequignot~1973; 
Shull \& van Steenberg~1982;
Landini \& Monsignori Fossi~1990; Landini \& Fossi~1991;
Pequignot, Petitjean, \& Boisson~1991; Arnaud \& Raymond~1992;
Verner et al.~1996),
dielectronic recombination (Aldrovandi \& Pequignot~1973;
Arnaud \& Raymond~1992;
Badnell et al.~2003, Badnell~2006;
Colgan et al.~2003, Colgan, Pindzola, \& Badnell~2004, 2005;
Zatsarinny et al.~2003, 2004a, 2004b, 2005a, 2005b, 2006;
Altun et al.~2004, 2005, 2006;
Mitnik \& Badnell~2004), and
neutralization and ionization by charge transfer reactions with
hydrogen and helium atoms and ions
(Kingdon \& Ferland fits\footnote{See:
http://www-cfadc.phy.ornl.gov/astro/ps/data/cx/hydro-\\gen/rates/ct.html},
based on Kingdon \& Ferland~1996,
Ferland et al.~1997, Clarke et al.~1998, Stancil et al.~1998;
Arnaud \& Rothenflug~1985). Our atomic data set is the
up-to-date compilation used in the steady-state
photoionization codes ION
(Netzer et al.~2005; and private communication) and Cloudy
(Ferland et al.~1998).

We assume that the gas is dust-free
(see below) and exclude neutralization processes by dust grains.  
Our code can also account for photoionization by a 
steady or time-varying external radiation field. 
However, in the computations presented here we exclude such radiation.

The time-dependent equations for the ion abundance
fractions, $x_i$, of element $m$ in ionization stage $i$ are,
\begin{equation}
\label{ion-eq}
\begin{array}{l}
d{x_i}/dt = x_{i-1}~~[q_{i-1}n_{\rm e} + \Gamma_{i-1}
+ k^{\rm H}_{\uparrow i-1}n_{\rm H^+}\\
\;\;\;\;\;\;\;\;\;\;\;\;\;\;\;\;\;\;\;\;\;\;\;\;\;\;+ k^{\rm He}_{\uparrow i-1}n_{\rm He^+}]\\
\;\;\;\;\;\;\;\;\;\;\; + x_{i+1}~~[\alpha_{i+1}n_{\rm e} +
k^{\rm H}_{\downarrow i+1}n_{\rm H^0}
+ k^{\rm He}_{\downarrow i+1}n_{\rm He^0}] \\
\;\;\;\;\;\;\;\;\;\;\; - x_{i}~~[(q_{i} + \alpha_{i})n_{\rm e} + \Gamma_i +
k^{\rm H}_{\downarrow i}n_{\rm H^0}
+ k^{\rm He}_{\downarrow i}n_{\rm He^0}\\
\;\;\;\;\;\;\;\;\;\;\;\;\;\;\;\;\;\;\;\;\; + k^{\rm H}_{\uparrow i}n_{\rm H^+}
+ k^{\rm He}_{\uparrow i}n_{\rm He^+}] \ \ \ .
\end{array}
\end{equation}
In this expression, $q_i$ and $\alpha_i$ are the rate coefficients
for collisional ionization and recombination (radiative plus
dielectronic), and
$k^{\rm H}_{\downarrow i}$, $k^{\rm He}_{\downarrow i}$,
$k^{\rm H}_{\uparrow i}$, and $k^{\rm He}_{\uparrow i}$
are the rate coefficients for charge transfer reactions
with hydrogen and helium that lead to ionization or neutralization.
The quantities
$n_{\rm H^0}$, $n_{\rm H^+}$, $n_{\rm He^0}$,
$n_{\rm He^+}$, and $n_{\rm e}$ are the particle densities (cm$^{-3}$)
for neutral hydrogen, ionized hydrogen, neutral helium, singly ionized helium, 
and electrons, respectively. $\Gamma_{i}$ are the photoionization rates of ions $i$,
due to externally incident radiation. Here, we set $\Gamma_i=0$. 

For each element $m$, the ion fractions 
$x_i\equiv n_{i,m}/(n_{\rm H} A_m)$
must at all times satisfy
\begin{equation}
\sum x_i = 1
\end{equation}
where $n_{i,m}$ is the density (cm$^{-3}$) of ions in ionization stage $i$
of element $m$,
$n_{\rm H}$ is total density of hydrogen nuclei, and $A_m$ is the
abundance of element $m$ relative to hydrogen. 
The sum is over all ionization stages of the element.

In our time-dependent code 
we can explicitly account for the effects of trapping of hydrogen and
helium recombination radiation
by switching appropriately from ``case A'' to ``case B'' recombination,
depending on the total cloud column density and temperature.
As we discuss in the Appendix, we find that such trapping has generally 
only a small effect on the ionization states of the metals
in the radiatively cooling gas.  We therefore present
results assuming optically thin case A conditions throughout.
We also neglect reabsorption of line and continuum radiation
emitted by metals in the cooling gas.
In the Appendix we provide computational estimates of the 
maximal column densities for which reabsorption of the cooling radiation
may be ignored. 

\subsection{Cooling}
\label{physics-cool}
The ionization equations~(\ref{ion-eq}) are coupled to an
energy equation for the time-dependent heating and cooling,
and resulting temperature variation.
Here we are interested in clouds undergoing pure radiative
cooling with no heat sources.

For an ideal gas, the gas pressure $P=Nk_{\rm B}T/V$, and the 
internal thermal energy $U=3/2~Nk_{\rm B}T$, where $T$ is the gas
temperature, $V$ is the total volume,
$N$ is the total number of particles in the system,
and $k_{\rm B}$ is the Boltzmann constant.
If $dQ$ is an
amount of heat lost (or gained) by the thermal electron gas, then
\begin{equation}\label{dq}
dQ = dU + PdV = \left(\frac{3}{2}+s\right) (Nk_{\rm B}dT+k_{\rm B}TdN).
\end{equation}
where
\begin{equation}
s = \left\{ \begin{array}{lcl}
0&,& {\rm isochoric~cooling,~~{\it dV=}\ 0} \\
1&,& {\rm isobaric~cooling,~~{\it dP=}\ 0} \\
\end{array}\right.
\end{equation}
For pure radiative cooling, with no heating, 
$dQ\equiv - n_e n_H \Lambda(T,x_i,Z) V dt$
where $\Lambda(T,x_i,Z)$ is the electron 
cooling efficiency (erg s$^{-1}$ cm$^3$),
which depends on the gas temperature, the ionization state, and
the total abundances of the heavy elements specified by the metallicity $Z$.
The electron cooling efficiency includes the removal of electron
kinetic energy via recombinations with ions, collisional ionizations,
collisional excitations followed by prompt line emissions,
and thermal bremsstrahlung.
For the low density  
``coronal limit'' that we consider,
$\Lambda$ is independent of the gas density or pressure.

In equation~(\ref{dq}) we do not include the ionization
potential energies as part of the total internal energy
(e.g.~Schmutzler \& Tscharnuter 1993) since $dQ$ refers
to the heat lost by the thermal electrons. In our
definition of $\Lambda$, therefore, the ionization
potential energy that is released as recombination radiation does not appear.
Only the kinetic energy of the recombining
electrons contributes to the cooling efficiency\footnote{The 
recombination radiation energy
$\epsilon_{\rm r}=\epsilon_{\rm KE} + \epsilon_{\rm I}$,
where $\epsilon_{\rm KE}$ is the thermal kinetic energy
of the recombining electron, and $\epsilon_{\rm I}$ is the
ionization energy of the recombined ion.
In our $\Lambda$ we include only $\epsilon_{\rm KE}$.}.
On the other hand, kinetic energy
removed via collisional ionization is included in our $\Lambda$.
If ionization potential energy {\it is} considered
as part of the total internal energy, then collisional
ionization does not lead to a net energy loss, since
the kinetic energy removed is merely stored as potential energy.
Either way of accounting for the energy losses leads to
the same temperature versus time relation $T(t)$.

It follows from equation~(\ref{dq}) that
the gas temperature declines at a rate given by (e.g., Kafatos 1973),
\begin{equation}
\label{dTdt}
\frac{dT}{dt} = - \frac{n_{\rm e} n_{\rm H} \Lambda(T,x_i,Z)}
{(3/2 +s)nk_{\rm B}} 
- \frac{T}{N} \frac{dN}{dt}
\end{equation}
where $n\equiv N/V$ is the total particle density of the gas.
Because of the $PdV$ work, the temperature declines more slowly 
for isobaric ($s=1$) cooling compared to isochoric ($s=0$) cooling.

The second term in equation~(\ref{dTdt}) reflects the relative change in the
total number of particles in the system. At temperatures $\gtrsim 10^4$ K,
where the hydrogen and helium remain ionized, the number of particles
remains approximately constant and this term may be neglected.
It plays a role at lower temperatures as the hydrogen recombines (Kafatos 1973).
For a primordial helium abundance, 
$A_{\rm He}\equiv n_{\rm He}/n_{\rm H} = 1/12$ (Ballantyne, Ferland, \& Martin~2000),
the temperature declines at a rate
\begin{equation}
\label{approxdT}
\frac{dT}{dt} 
\simeq
\frac{n \Lambda}{4.34\times (3/2+s)k_{\rm B}},
\end{equation}
for gas in which the hydrogen and helium are fully ionized.

Equation~(\ref{dTdt}) can also be expressed as (e.g., Shapiro \& Moore~1976),
\begin{equation}
\label{SMdTdt}
\frac{3}{2} \frac{dP}{dt} -
\frac{5}{2} \frac{P}{\rho} \frac{d\rho}{dt} = 
- n_{\rm e} n_{\rm H} \Lambda(T,x_i,Z)
\end{equation}
where $\rho = \mu n$ is the mass density
of the gas, and $\mu$ is the mean mass per particle.
For hydrogen and helium fully ionized,
and a primordial helium abundance,
$\mu=16/27~m_{\rm H}$ where $m_{\rm H}$ is the proton mass.

Equations~(\ref{ion-eq}) and (\ref{dTdt}) imply
that in the absence of external radiation ($\Gamma_i=0$)
the derivatives of the ion fractions with respect to temperature, 
$dx_i/dT \equiv (dx_i/dT)(dT/dt)$,
are independent of the gas density or pressure.
Because collisional ionization, electron-ion
recombination, hydrogen and helium charge-transfer reactions,
and the gas cooling processes are all
``two-body'' interactions, with rates per unit volume
proportional to $n^2$, the density dependence divides out.
The solutions for the ion fractions as functions of
the instantaneous gas temperature, $x_i(T)$, are therefore
independent of the assumed gas density or pressure.

Given a set of non-equilibrium ion abundances, $x_i(T)$,
and an assumed metallicity $Z$, we use the cooling functions
included in Cloudy (version 06.02, Ferland et al.~1998) 
to calculate $\Lambda(T,x_i,Z)$.
We present results for
$Z$ ranging from $10^{-3}$ to 2 times solar.
For $Z=1$ we adopt the elemental abundances for
C, N, O, Mg, Si, S, and Fe reported by
Asplund et al.~(2005a) for the photosphere of the Sun. 
An exception is Ne, for which
we adopt the ``enhanced'' abundance recommended by Drake \& 
Testa~(2005)\footnote{An enhanced Ne abundance may reconcile the
``new'' Asplund photospheric abundances with solar models and
helioseismology (Bahcall, Basu, \& Serenelli~2005; Anita \& Basu 2005;
but see Asplund et al.~2005b; Schmelz et al.~2005).
See, however, Ayres et al.~(2006).}. 
We list our assumed solar abundances in Table \ref{solar}.
In all computations we assume a primordial helium abundance
$A_{\rm He}=1/12$, independent of $Z$.
The elements we include are the dominant coolants over the entire temperature
range we study. Cooling by other elements is negligible.

\begin{deluxetable}{lr}
\tablewidth{0pt}
\tablecaption{Solar Elemental Abundances}
\tablehead{
\colhead{Element} & 
\colhead{Abundance}\\
\colhead{} &
\colhead{(X/H)$_{\odot}$} }
\startdata
Carbon   & $2.45\times10^{-4}$ \\
Nitrogen & $6.03\times10^{-5}$ \\
Oxygen   & $4.57\times10^{-4}$ \\
Neon     & $1.95\times10^{-4}$ \\
Magnesium& $3.39\times10^{-5}$ \\
Silicon  & $3.24\times10^{-5}$ \\
Sulfur   & $1.38\times10^{-5}$ \\
Iron     & $2.82\times10^{-5}$ \\
\enddata
\label{solar}
\end{deluxetable}

Non-equilibrium effects become important when the cooling time
\begin{equation}
\label{tcool}
t_{{\rm c}} \equiv \frac{T}{|dT/dt|} \simeq \frac{4.34\times (3/2+s) k_{\rm 
B} T}{n \Lambda(T,x_i,Z)}
\end{equation}
becomes short compared to the recombination time for an ion $i$,
\begin{equation}
\label{trec}
t_{{\rm r},i} \simeq \frac{1}{n \alpha_{i}^{\rm tot}} \ \ \ ,
\end{equation}
where $\alpha_i^{\rm tot}$ is the total recombination 
coefficient at the temperature $T$.
The ionization state remains close to 
``collisional ionization equilibrium'' (CIE) when
$t_{\rm c}/t_{\rm r,i}>>1$ for the most abundant ions at the given temperature.
Non-equilibrium effects become significant when $t_{\rm c}/t_{\rm r,i} < 1$.

Because different ions have different recombination times,
the recombination-lag is not similar for all ions.
Ions with longer recombination times, in particular
the helium-like ions, may be expected to
persist over a wider range of temperatures when out of equilibrium.

Crucially, the ratios $t_{\rm c}/t_{\rm r,i}$ are 
{\it independent} of the gas density or pressure, so that departures
from CIE are independent of the assumed gas density or pressure,
for both isochoric or isobaric cooling. However, because
$t_{\rm c}$ is shorter for isochoric cooling,
non-equilibrium effects may be expected to be somewhat 
more pronounced for isochoric cooling compared
to isobaric cooling.

The primary, and essentially only, ``free parameter'' in the problem
is the metallicity $Z$ of the gas, through its control of $\Lambda$
and the cooling time. A higher $Z$ leads to enhanced metal line
cooling, and larger departures from CIE, and vice versa.

In our computations we assume that the gas is dust-free,
and we do not consider cooling via gas-grain collisions,
a process that can dominate the total cooling at high temperatures
if a high dust mass can be maintained
(Ostriker \& Silk 1973, Draine 1981). 
The neglect of dust is appropriate when the dust sputtering 
destruction time scale, 
$t_{\rm s}$, is shorter than the initial cooling rate, $t_{\rm c}$, so
that any initial dust is destroyed before appreciable
gas cooling occurs.  Here we rely on the computations
of Smith et al.~(1996) who found that $t_{\rm s}/t_{\rm c} < 1$
for $T\gtrsim 3\times 10^6$ K (see their Fig.~7).  
In our calculations we consider gas cooling from
initial temperatures greater than this, and
we assume that the dust is instantaneously destroyed. The gas 
then evolves with constant gas-phase elemental abundances
specified by the assumed initial metallicity $Z$. 

\subsection{Numerical Procedure}
\label{numerical}
For the atomic elements that we include,
equations~(\ref{ion-eq}) and (\ref{dTdt}) are a set of $103$ coupled
ordinary differential equations (ODEs).  
In our numerical procedure we advance the isochoric solutions 
in small pressure steps $\Delta P = \varepsilon~P$
where $\varepsilon \lesssim 0.05$,
and $P$ is the current pressure, associated with a temperature $T(P)$.
For isobaric conditions, we advance the solutions in small density
steps $\Delta \rho = \varepsilon~\rho$, where $\rho$ is the current 
mass density, associated with a temperature $T(\rho)$.

For any $T$ 
we compute the total cooling
rate by passing the current (non-equilibrium) ionization fractions
$x_i(T)$ to the Cloudy cooling functions.
We then use equation~(\ref{SMdTdt}) to compute the time interval
$\Delta t$ associated with pressure change $\Delta P$ (or density change
$\Delta\rho$).
We then integrate equations~(\ref{ion-eq}) over the time interval $\Delta t$
using a Livermore ODE solver\footnote{
This package integrates initial value problems for stiff or non-stiff systems,
and switches automatically between stiff and non-stiff methods
as necessary.
See: http://www.netlib.org/odepack/} 
(Hindmarsh~1983).
When integrating the ionization equations~(\ref{ion-eq}),
we assume that over the time step $\Delta t$, the 
pressure (or density) evolves linearly with time.
In the integrations, the estimated local errors on the fractional abundances
were controlled so as to be smaller than $10^{-6}$, $10^{-5}$, and $10^{-4}$
for hydrogen, helium, and metal-ions, respectively.

The above procedure is repeated with the new values of the
ionization fractions,
$x_i(T+\Delta T)$,
down to a minimum temperature $T_{\rm low}$.
Here we set $T_{\rm low}=10^4$~K.
We verify convergence by re-running the
computation at higher resolution (smaller $\varepsilon$), and confirming
that the resulting ionization and cooling rates as functions of time
remain unaltered.

\section{Isochoric Versus Isobaric Cooling}
\label{isowhat}

A 
gas cloud will cool isochorically (i.e., at constant mass-density and volume)
when the cooling time as defined by equation~(\ref{tcool}) is
short compared to the dynamical time
\begin{equation}
t_d \equiv \frac{D}{c_s}
\end{equation}
where $D$ is the cloud diameter, and $c_s=\sqrt{k_{\rm B}T/\mu}$
is the sound speed. 
Cooling is isobaric (i.e., occurs at constant pressure
and decreasing volume) when the cooling time is long compared to
the dynamical time. 

\begin{figure*}
\epsscale{1}
\plottwo{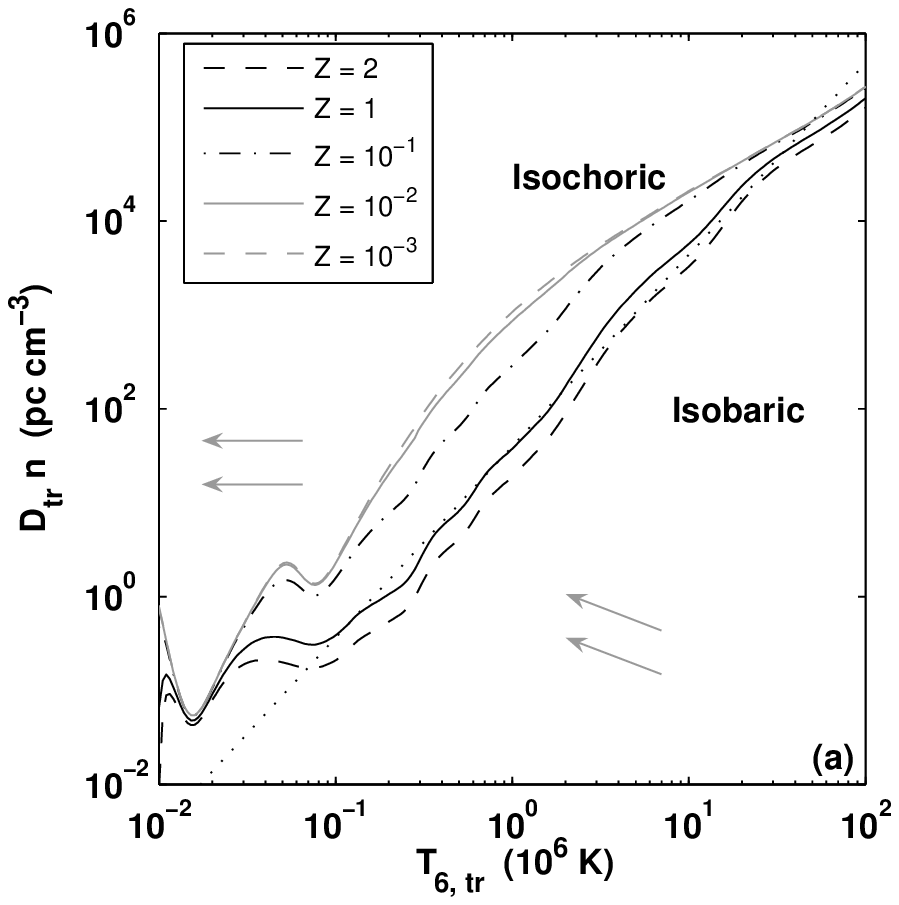}{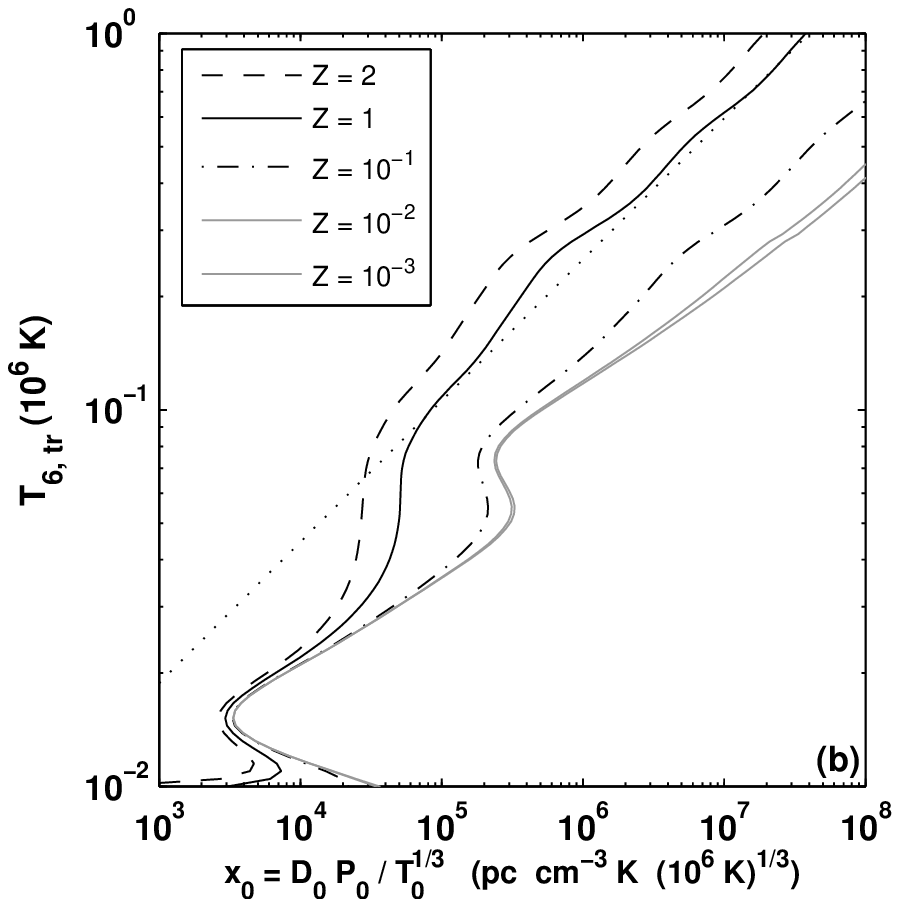}
\caption{Critical column densities, $D_{\rm tr}n$, and temperatures, 
$T_{\rm tr}$,
for the transition from isobaric to isochoric cooling, for clouds with
metallicities $Z$ ranging from $10^{-3}$ to $2$ times solar.
(a) Transition column, $D_{\rm tr}n$, versus 
transition temperature, $T_{\rm tr}$, as given by equation~(\ref{C2B}).
The dotted line is the power-law approximation for $Z=1$.
In the region above the curves the cooling is isochoric. Below the
curves the cooling is isobaric.
The arrows indicate the direction in which the gas evolves as it cools.
(b) Transition temperature, $T_{\rm tr}$, 
from isobaric to isochoric cooling, versus the initial state parameter
$x_0\equiv D_0P_0/T_0^{1/3}$, as given by equation~(\ref{B2C}).
The dotted line is the power-law approximation for $Z=1$.
}
\label{choose-case}
\end{figure*}

The dynamical evolution is therefore determined by the ratio
\begin{equation}
\label{tc_over_td}
\frac{t_{\rm c}}{t_d} \simeq 4.34 \times \frac{(3/2+s) (k_{\rm 
B}T)^{3/2}}{\mu^{1/2} n \Lambda D} \ \ \ .
\end{equation}
Cooling is isochoric when $t_c/t_d << 1$ and 
is isobaric when $t_c/t_d >> 1$.
Equation~(\ref{tc_over_td}) shows that for a given cloud size
and density the cooling
is isobaric if the temperature is sufficiently high. This is
because $T^{3/2}/\Lambda$ generally increases rapidly with
increasing temperature (see \S 5). Conversely, for a given temperature
and density, the cooling is isochoric for sufficiently large $D$.

For isochoric cooling, $D$ and $n$ remain constant
with time. For these conditions
$t_{\rm c}/t_d \propto T^{3/2}/ \Lambda$,
which generally decreases as the gas cools.
Hence, cooling that is initially isochoric remains isochoric.

For isobaric cooling, a spherical cloud with initial
diameter $D_0$, density $n_0$, and temperature
$T_0$, will contract to a size $D = D_0 (T/T_0)^{1/3}$ with density $n = n_0 
T_0/T$ after cooling to a temperature $T$.
Thus, as the cloud cools and contracts
$t_{\rm c}/t_d \propto T^{13/6}/\Lambda$. This ratio
decreases rapidly as the gas cools, so that a transition to isochoric
cooling must eventually occur.

The transition from isobaric to isochoric cooling occurs
at a critical cloud size, $D_{\rm tr}$,
and temperature, $T_{\rm tr}$, at which $t_{\rm c} \simeq t_d$.
It follows from 
equation~(\ref{tc_over_td}) that the transition occurs at a critical column
density 
\begin{equation}
\label{C2B}
\begin{array}{l}
D_{\rm tr}n \cong 10~k_{\rm B}^{3/2} \mu^{-1/2}
\frac{T_{\rm tr}^{3/2}}{\Lambda(T_{\rm tr})}\\
\;\;\;\;\;\;\;\stackrel{_{(Z=1)}}{\cong}
 40 \times T_{6,{\rm tr}}^{2.04}~{\rm pc}~{\rm cm}^{-3}
\end{array}
\end{equation}
where $n$ is the gas density.
Given that non-equilibrium effects reduce the
cooling efficiencies by only factors a few (see \S 5) compared to CIE cooling,
we use the CIE cooling efficiencies, $\Lambda_{\rm eq}$,
in equation~(\ref{C2B}) to estimate $D_{\rm tr}$ and $T_{\rm tr}$.
The resulting estimates for the critical sizes and temperatures are then
independent of initial conditions. 
For the numerical evaluation in equation~(12) we use
the power-law approximation
$\Lambda_{\rm eq}^{Z=1} = 2.3\times10^{-19}~T^{-0.54}$~erg~cm$^3$~s$^{-1}$
for equilibrium cooling at solar metallicity (see \S 5). 

For a cloud that initially cools isobarically at pressure $P_0$, from
an initial diameter $D_0$ and temperature $T_0$, 
it then follows that 
\begin{equation}
\label{B2C}
\begin{array}{l}
\frac{D_0 P_0}{T_0^{1/3}} \cong 10~k_{\rm B}^{5/2}~\mu^{-1/2}
\frac{T_{\rm tr}^{13/6}}{\Lambda(t_{\rm tr})}\\
\;\;\;\;\;\;\;\stackrel{_{(Z=1)}}{\cong}
4\times10^7~T_{6,{\rm tr}}^{2.7}~\frac{{\rm pc}~({\rm cm}^{-3}~{\rm K})}{(10^6~{\rm K})^{1/3}}
\end{array}
\end{equation}
which provides an implicit relation for the temperature,
$T_{\rm tr}$, at which the transition to isochoric cooling occurs
(cf.~Edgar \& Chevalier 1986). The numerical evaluation is
for the power-law cooling efficiency at solar metallicity. 

In Figure \ref{choose-case} we plot $D_{\rm tr}n$ versus $T_{\rm tr}$
(left panel), and $T_{\rm tr}$ versus 
the parameter $x_0\equiv D_0P_0/T_0^{1/3}$ (right panel)
as given by equations~(\ref{C2B}) and (\ref{B2C}). We present results for
metallicities $Z$ ranging
from $10^{-3}$ to $2$ times solar, assuming 
the CIE cooling efficiencies we compute in \S 5.

In Figure \ref{choose-case}a, the cooling for each $Z$ is isochoric
for temperatures and cloud column densities
to the left and above of the curves.
The cooling is isobaric to the right and below the curves.
The dotted line in Figure \ref{choose-case}a
is the power-law approximation given in equation~(\ref{C2B}). 
The arrows in Figure \ref{choose-case}a
indicate the direction in which the gas evolves as it cools.
For isochoric cooling the column $Dn$ remains constant, and the gas
cools along horizontal trajectories.
For isobaric cooling $Dn \propto T^{-2/3}$,
as indicated by the inclined arrows,
showing that the gas must eventually make the transition to 
isochoric cooling. For example, a condensation
with diameter $D=10$~kpc, density $n=10^{-5}$~cm$^3$, and temperature
$T=10^6$~K (corresponding to a thermal pressure $P=10$~cm$^{-3}$~K)
cools isobarically, since $Dn=0.1$~pc~cm$^{-3}$, whereas for this
temperature $D_{\rm tr}n=40$~pc~cm$^{-3}$
for $Z=1$, or $10^3$~pc~cm$^{-3}$ for $Z<10^{-2}$.

In Figure \ref{choose-case}b, $T_{\rm tr}$ is the temperature at 
which the transition from isobaric to isochoric cooling occurs, 
given an initial state defined by the parameter $x_0$.
In computing $T_{\rm tr}$ we again use the equilibrium 
cooling efficiencies for the different values of $Z$.
The dotted line in Figure \ref{choose-case}b shows the power-law
approximation for $Z=1$ as given in equation~(\ref{B2C}).
For our above example (with $D=10$~kpc, $P=10$~cm$^3$~K, $T=10^6$~K), 
the parameter $x_0=10^5$, so that 
the initial isobaric condensation will begin cooling isochorically
at $T_{\rm tr}=10^5$~K for $Z=1$, or at $4\times10^4$~K for $Z < 10^{-2}$.

As we show in the Appendix, a cooling gas cloud remains
optically thin up to column densities
that are much greater than the critical transition columns
in Figure \ref{choose-case}. Expressions~(\ref{C2B}) and (\ref{B2C}),
and the curves in Figure \ref{choose-case}
are therefore of broad applicability for determining the
isobaric versus isochoric dynamical evolution for radiatively cooling clouds.

\section{Ionization Fractions}
\label{ionization-res}
We have carried out computations of the collisionally controlled
ionization states of the elements H, He, C, N, O, Ne, Mg, Si, S, and Fe, 
as functions of gas temperature for three sets of assumptions. First we assume
CIE imposed at all $T$. Then we consider the non-equilibrium 
ionization states as functions of the 
time-dependent temperature for radiatively cooling gas, for constant
pressure and constant density evolutions.
Our results are displayed in Figure \ref{ion-res-fig} 
and listed in tabular form
in Tables 2-4, and in additional tables and figures that we describe below.

The left-hand panels of Figure \ref{ion-res-fig} show our results for the
CIE ion fractions $x_i^{\rm eq}(T)$. Each panel displays
the ionization stages of a particular element.
In our CIE computations we have integrated the time-dependent
equations~(\ref{ion-eq}) to the equilibrium states
reached at late times when the formation and
destruction rates for each ion are equal.
We performed these time integrations at fixed 
temperatures $T$ ranging from 10$^4$ to 10$^8$ K.
For purposes of comparison with our non-equilibrium
computations, in Figure \ref{ion-res-fig} we display results for
temperatures between $10^4$ and $5\times 10^6$ K only. 
Results for higher temperatures, up to 10$^8$~K where nearly all of the
ions become fully stripped, are given in Table 2.
We verified that the CIE solutions we obtained by time-integration
are identical to those found by solving the set of 
algebraic equations for $x_i$ 
obtained by setting $dx_i/dt=0$ in equation~(\ref{ion-eq}), 
as appropriate for steady state\footnote{We solved the algebraic 
equations using both
Cloudy (Ferland et al.~1998) and ION (Netzer et al.~2004)
assuming identical input atomic data sets. These two
``photoionization codes'' assume steady-state conditions.}.

\begin{deluxetable}{lllll}
\tablewidth{0pt}
\tablecaption{CIE~Ion Fractions}
\tablehead{
\colhead{Temperature} & 
\colhead{H$^0$/H} & 
\colhead{H$^+$/H} & 
\colhead{He$^0$/He} & 
\colhead{\ldots}\\
\colhead{(K)} &
\colhead{} &
\colhead{} &
\colhead{} &
\colhead{} }
\startdata
$1.00\times10^4$ & $9.99\times10^{-1}$ & $9.30\times10^{-4}$ & $1.00\times10^0$ & \ldots \\
$1.05\times10^4$ & $9.97\times10^{-1}$ & $2.88\times10^{-3}$ & $1.00\times10^0$ & \ldots \\
$1.10\times10^4$ & $9.93\times10^{-1}$ & $6.98\times10^{-3}$ & $1.00\times10^0$ & \ldots \\
$1.15\times10^4$ & $9.84\times10^{-1}$ & $1.51\times10^{-2}$ & $1.00\times10^0$ & \ldots \\
$1.20\times10^4$ & $9.69\times10^{-1}$ & $3.05\times10^{-2}$ & $1.00\times10^0$ & \ldots \\
\enddata
\tablecomments{
(1) The complete version of this table is in the electronic edition of
the Journal.  The printed edition contains only a sample. 
(2) CIE ion fractions are given for $Z=1$. For $T\lesssim10^4$~K 
the electron density, and associated ion fractions, depend on $Z$.}
\end{deluxetable}

For an ionized hydrogen gas,
the CIE ion fractions are independent of the
metallicity $Z$, and are ``universal'' functions of
the gas temperature $T$, as determined by the microscopic
steady-state balance between electron impact collisional
ionization, electron-ion recombination, and charge transfer\footnote{For 
$T\lesssim 10^4$~K, hydrogen
becomes neutral and the electron density depends
on the assumed metallicity. At these low temperatures
the associated CIE metal ion fractions then do vary with $Z$.}.
Our CIE solutions, and in particular the shapes and peak
positions of the $x_i^{\rm eq}(T)$ curves,
are in excellent agreement with previous computations 
available in the literature (see \S 1).
Our CIE ion fractions differ only slightly from 
the widely quoted results of
Sutherland \& Dopita (1993). For example,
as found by Sutherland \& Dopita for carbon (see their Fig.~3),
$x({\rm C^{3+}})$ peaks at a value of $0.3$ around a narrow
temperature range close to $10^5$ K, whereas $x({\rm C^{4+}})$
is close to 1 for a broad temperature range from 
$\sim1.5\times10^5$~K to $\sim7\times10^5$~K. 
The widths of the $x_i^{\rm eq}(T)$ curves
reflect the magnitudes of the ionization potentials
of the ions. For example,
C$^{4+}$, N$^{5+}$, and O$^{6+}$ persist for a wide range of
temperatures due to the increase in
thermal energy scale that is  
required to remove the inner K-shell electrons.

The right-hand panels of Figure \ref{ion-res-fig} display our results
for non-equilibrium ionization in radiatively
cooling gas for solar ($Z=1$) composition.
For time-dependent cooling the behavior depends on the initial
conditions. Here we assume that
the gas is cooling from an initially
hot, $T>5\times 10^6$ K, equilibrium state.
For such high temperatures the cooling times are long,
and there is generally sufficient time for the gas
to reach CIE before appreciable cooling occurs.
Our results
are also listed in ``on-line'' Tables~3 and 4. Results for
other metallicities are presented in additional electronic on-line
figures and tables. In each panel in Figure \ref{ion-res-fig}, 
the black curves
are for isochoric cooling, and the light grey curves are
for isobaric cooling. Departures from
CIE become apparent for $T\lesssim 10^6$ K,
for which cooling becomes rapid compared to recombination.

When departures from equilibrium occur,
the gas at any temperature
remains over-ionized compared to CIE, as recombination lags behind cooling.  
The recombination lag affects all elements including hydrogen and helium.
For example, in CIE the hydrogen is half-ionized 
at $T=1.6\times 10^4$~K, and is fully neutral ($99.9\%$) at $10^4$~K.
However, for non-equilibrium (isochoric) cooling for $Z=1$ 
the hydrogen is only $9\%$ neutral at $10^4$~K.
The metals are affected in a similar manner, with
$x_i(T)$ remaining 
overionized compared to CIE
at low temperatures. 
The recombination lags are largest for ions with
small recombination efficiencies. Thus, He-like ions,
such as C$^{4+}$, N$^{5+}$, and O$^{6+}$, which have
long recombination times, persist to much lower temperatures
in radiatively cooling gas compared to CIE.
For example, for solar composition $x({\rm C^{4+}}) > 0.2$
down to $T\sim1.4\times10^4$~K for isochoric cooling,
and $2.3\times10^5$~K for isobaric cooling, whereas in CIE
$x^{\rm eq}({\rm C^{4+}})$ becomes vanishingly small
for $T$ below $\sim8\times10^4$~K.
Similarly, the non-equilibrium abundance of the
widely observed ion O$^{5+}$ remains greater than 10\% of
its peak value of $0.11$ occurring at $3.0\times10^5$~K, 
down to temperatures of $2.0\times10^4$~K for isochoric cooling,
and $1.0\times10^5$~K for isobaric cooling.
In CIE, O$^{5+}$ vanishes below $\sim 2\times10^5$~K.    

Figure \ref{ion-res-fig} 
shows that at high temperatures, where CIE is attained, the isobaric and
isochoric $x_i(T)$ curves converge as they must. However, for
lower temperatures, the recombination lags are greater 
for isochoric cooling compared to isobaric cooling. 
This is due to the shorter isochoric cooling times
(see equation~[\ref{tcool}]).  
The isochoric and isobaric ion fractions differ
most substantially for ions with the largest
recombination lags. The differences grow with increasing metallicity
as the cooling times become shorter.

Another important feature of the time-dependent ion distributions
is the ``double-peaked'' behavior of the $x_i(T)$ curves seen for many ions,
e.g. C$^{++}$, C$^{3+}$, Si$^{3+}$, S$^{4+}$, and S$^{5+}$.
This behavior reflects the temperature dependence of
dielectronic versus radiative recombination (Kafatos 1973).
For such ions,
dielectronic recombination is the dominant removal mechanism from the
high-temperature peaks to the central minima. 
At lower temperatures dielectronic recombination weakens, and
the ion fractions rise again due to
continuing recombination from the persisting higher ionization stages. 
As the temperature continues to decrease, radiative
recombination becomes increasingly effective and the ion fractions drop again,
leading to the low-temperature peaks. The double-peaked behavior
is a time-dependent effect, because
the low-temperature peaks occur at temperatures for which the
ions no longer exist in CIE.

Our non-equilibrium $x_i(T)$ curves are in qualitative
agreement with previous computations
(Kafatos~1973, Shapiro \& Moore~1976, Sutherland and Dopita~1993,
Schmutzler \& Tscharnuter~1993). However, there are some 
significant differences in detail which are due,
at least in part, to differences in the assumed atomic 
processes and rate coefficients. For example, the early studies by
Kafatos (1973) and Shapiro \& Moore (1976) did not include
charge transfer reactions with hydrogen or helium,
which affect the ion fractions
for $T\lesssim 10^5$~K. Differences between our results and the
more recent computations of Sutherland and Dopita (1993) and
Schmutzler \& Tscharnuter (1993) are likely due to 
differences in dielectronic rate coefficients (which control
the ``double-peak'' behavior)
and in the assumed solar abundances. For example,
Sutherland \& Dopita adopted the Anders \& Grevesse (1989)
abundances with ${\rm O/H}_\odot = 8.5\times10^{-4}$,
whereas we adopt the more recent
Asplund et al.~(2005a) value of $4.6\times10^{-4}$ (see Table \ref{solar}).
Different assumed abundances lead to altered cooling times and
departures from CIE.

Our results for the non-equilibrium ion fractions, $x_i(T)$, for 
metallicities $Z$ equal to $10^{-3}$, $10^{-2}$, $10^{-1}$, and $2$, are
presented in electronic ``on-line'' Figures 3-6, and Tables 5-12.
Here we consider two examples of
how the assumed metallicity affects the ion fractions.
In Figure \ref{OVI_Z} we display the 
O$^{5+}$ and O$^{++}$ distributions for the different values of $Z$
in isochorically cooling gas.
At high temperatures ($\gtrsim 5\times 10^6$~K) 
the cooling times are long for all $Z$, and the O$^{5+}$ and O$^{++}$
abundances converge to the CIE distributions (thick grey curves).
Departures from CIE occur at lower temperatures. 
The departures are largest for large
$Z$ where the cooling times are shortest.
For example, for $Z=2$ the O$^{5+}$ distribution is very broad, and
this ion persists down to $\sim 10^4$~K, remaining within a factor $3$ 
of its peak abundance of $0.1$ at $T=1.7\times10^4$~K.
For smaller $Z$ the cooling times are longer,
and the deviations from CIE become smaller.

For very low metallicities ($Z\lesssim 10^{-2}$), 
the metals contribute negligibly to the gas cooling (see \S 5),
and the cooling times are independent of $Z$
(e.g.~Boehringer \& Hensler 1989).
In the ``low-Z'' limit, the ion-fraction
\begin{figure*}
\plotone{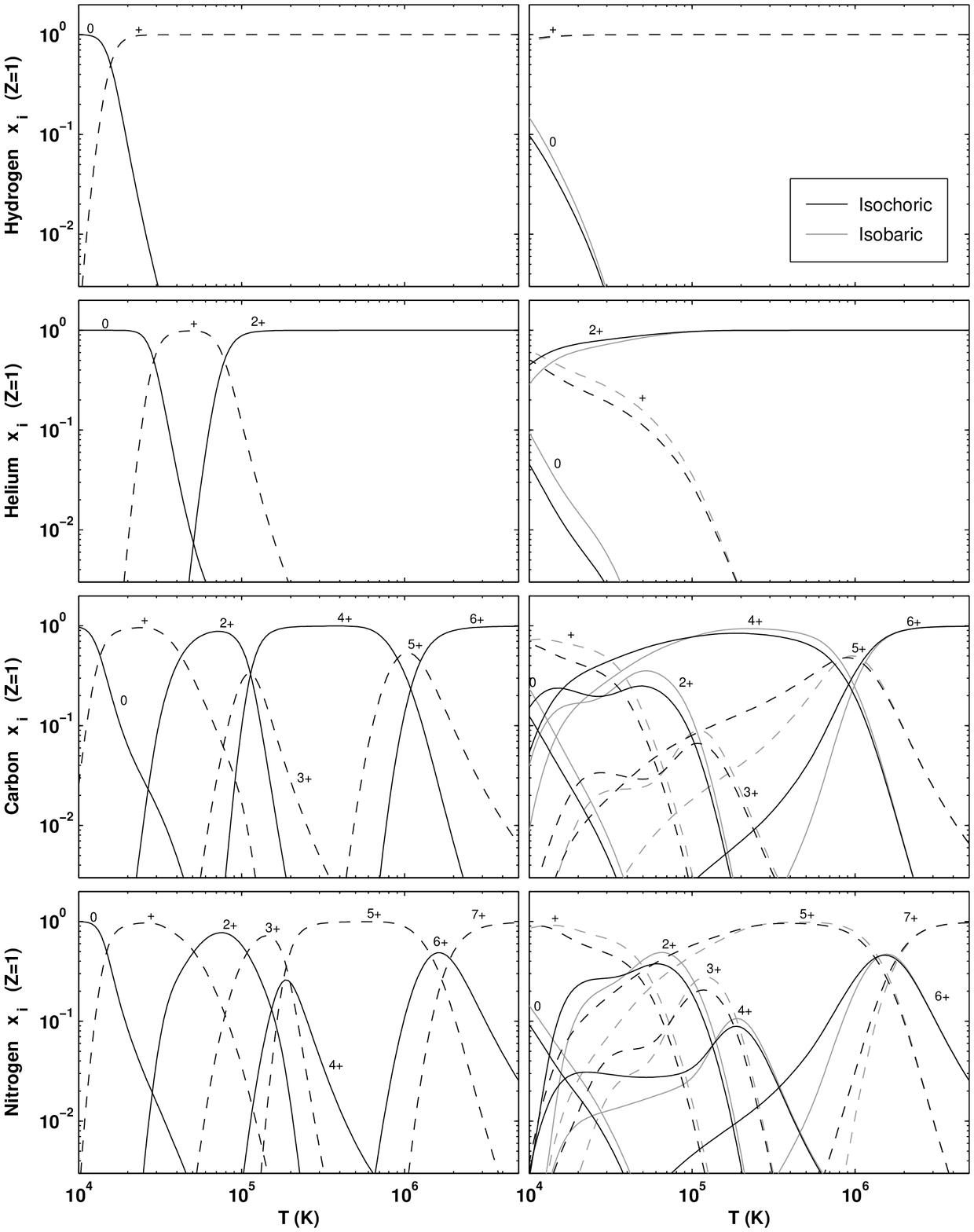}
\caption{
Ion fractions $x_i \equiv n_{i,m}/n_{\rm H}A_{\rm m}$,
versus gas temperature. Each row displays results for a 
different element. Left hand panels are for collisional
ionization equilibrium (CIE). Right hand panels are for
non-equilibrium isochoric and isobaric cooling for $Z=1$ 
times solar metallicity gas. Dark curves are
for isochoric cooling. Light curves are for isobaric cooling.
Panels show H, He, C, and N.}
\label{ion-res-fig}
\end{figure*}
\setcounter{figure}{1}
\begin{figure*}
\plotone{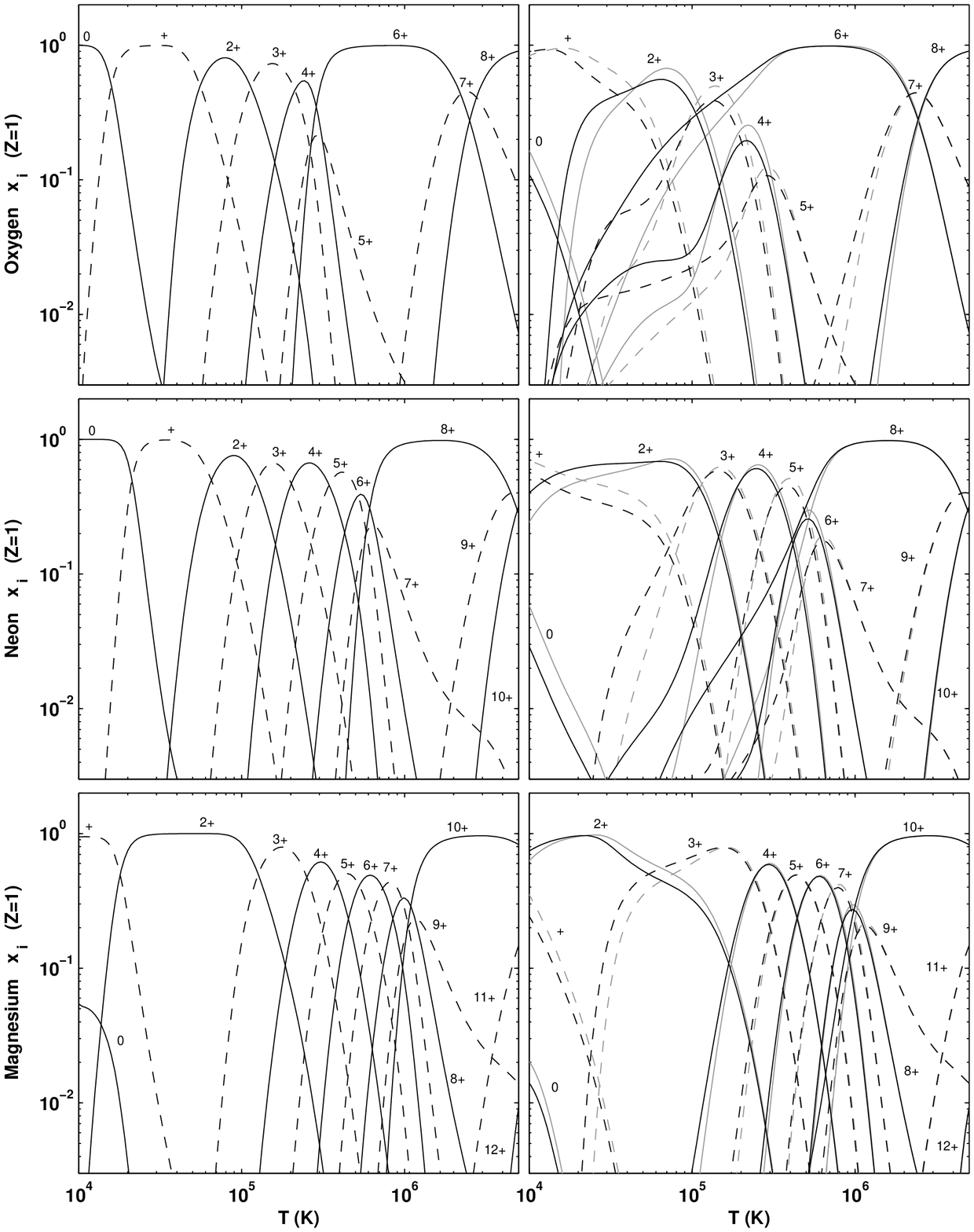}
\caption{Continued. Panels show O, Ne, and Mg.}
\end{figure*}
\setcounter{figure}{1}
\begin{figure*}
\plotone{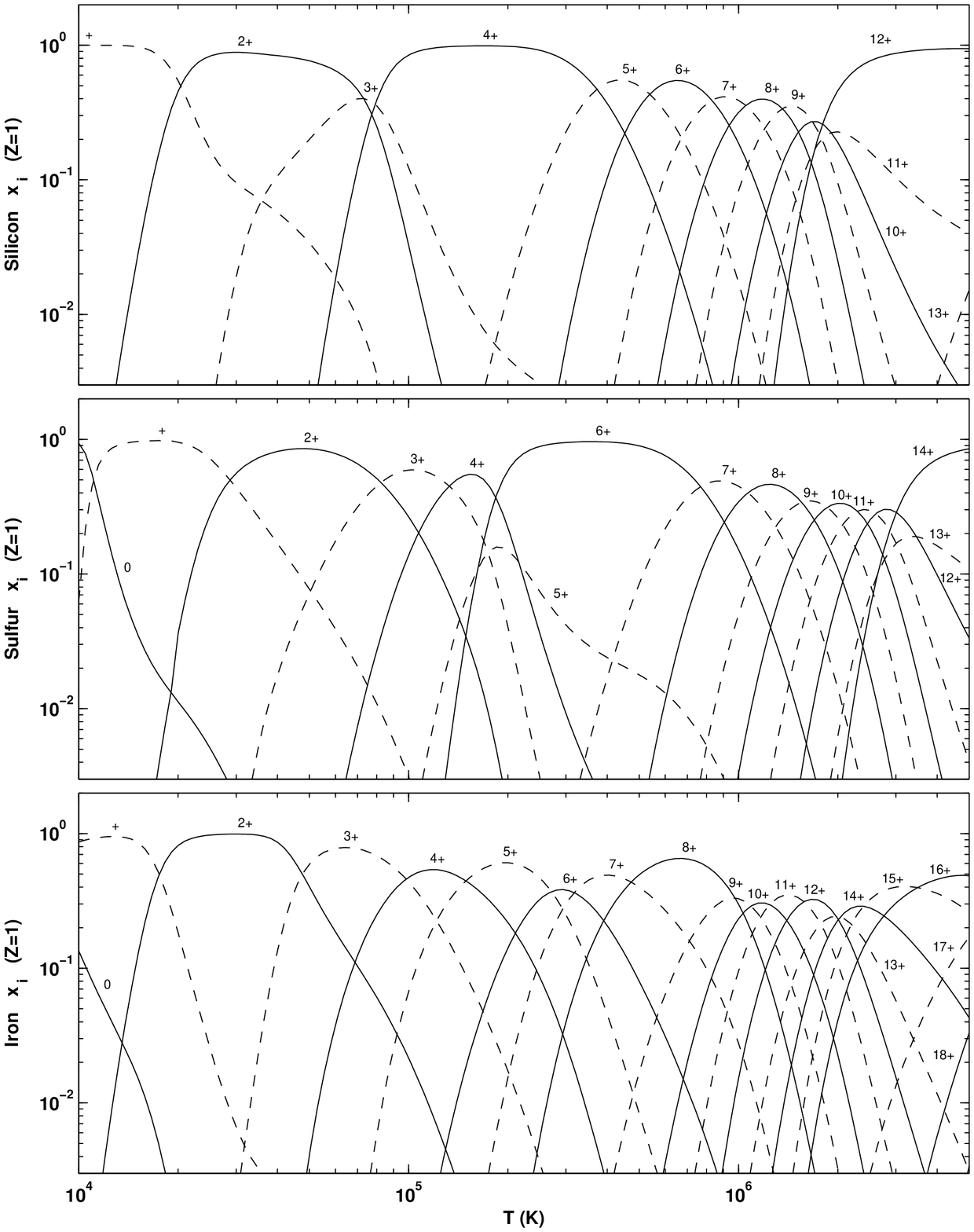}
\caption{Continued. Panels show CIE ion fractions for Si, S, and Fe.}
\end{figure*}
\setcounter{figure}{1}
\begin{figure*}
\plotone{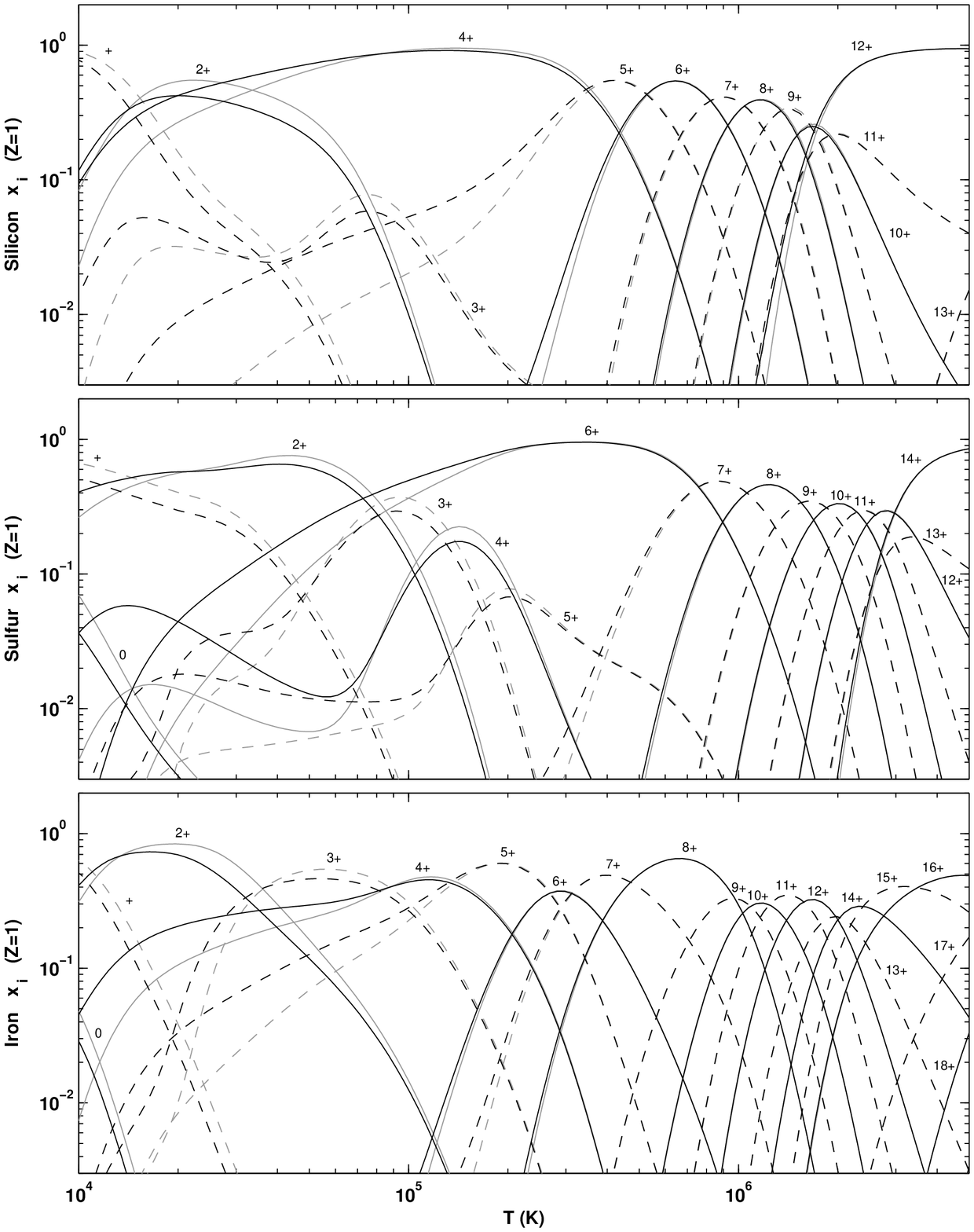}
\caption{Continued. Panels show time-dependent ion fractions for Si, S, and Fe.}
\end{figure*}
\clearpage
curves therefore 
converge to specific ``primordial'' forms, which may or may not
approach the CIE distributions. This depends on whether the low-$Z$
cooling time is long or short compared to the ion recombination time.
For example, Figure \ref{OVI_Z}a shows that at low $Z$, the O$^{5+}$ 
fractions approach CIE. However,
Figure \ref{OVI_Z}b shows that O$^{++}$ remains out of equilibrium
in the low-$Z$ limit. This is an example of an ion whose abundance
distribution can {\it never} reach equilibrium, 
for {\it any} $Z$, in radiatively cooling gas.
Such ions are generally those that appear at
$T\lesssim 2\times 10^5$~K, where the low-$Z$ cooling times remain
sufficiently short due to efficient cooling by hydrogen and
helium Ly$\alpha$ emissions (see \S 5).  

\setcounter{figure}{6}
\begin{figure*}
\epsscale{1}
\plottwo{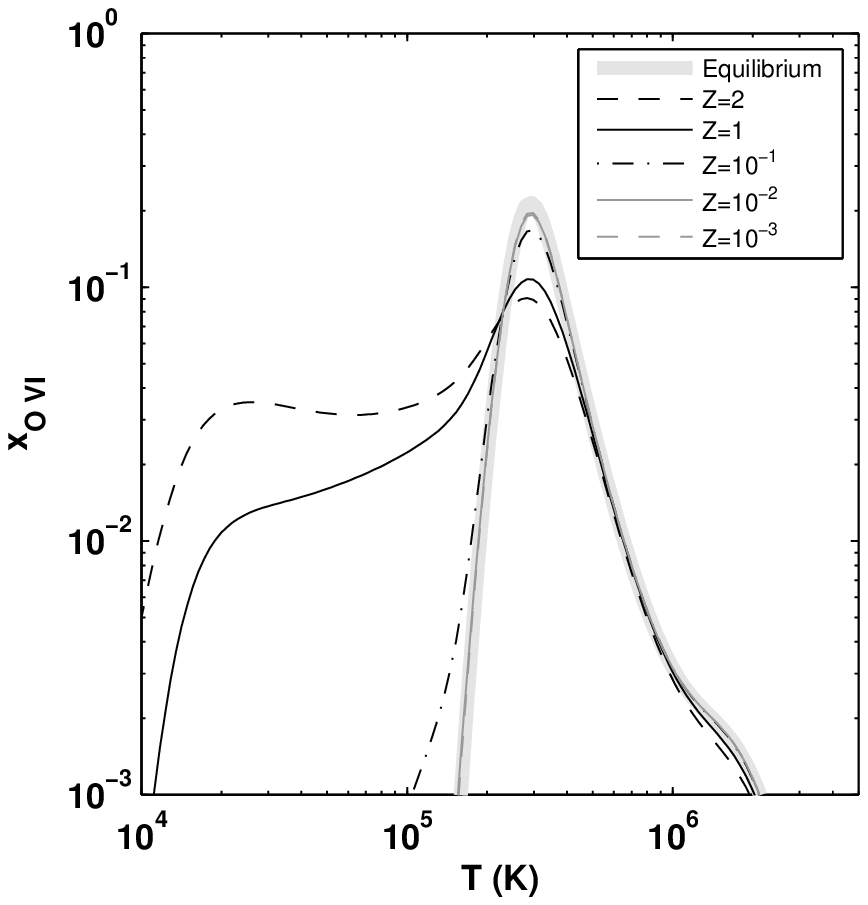}{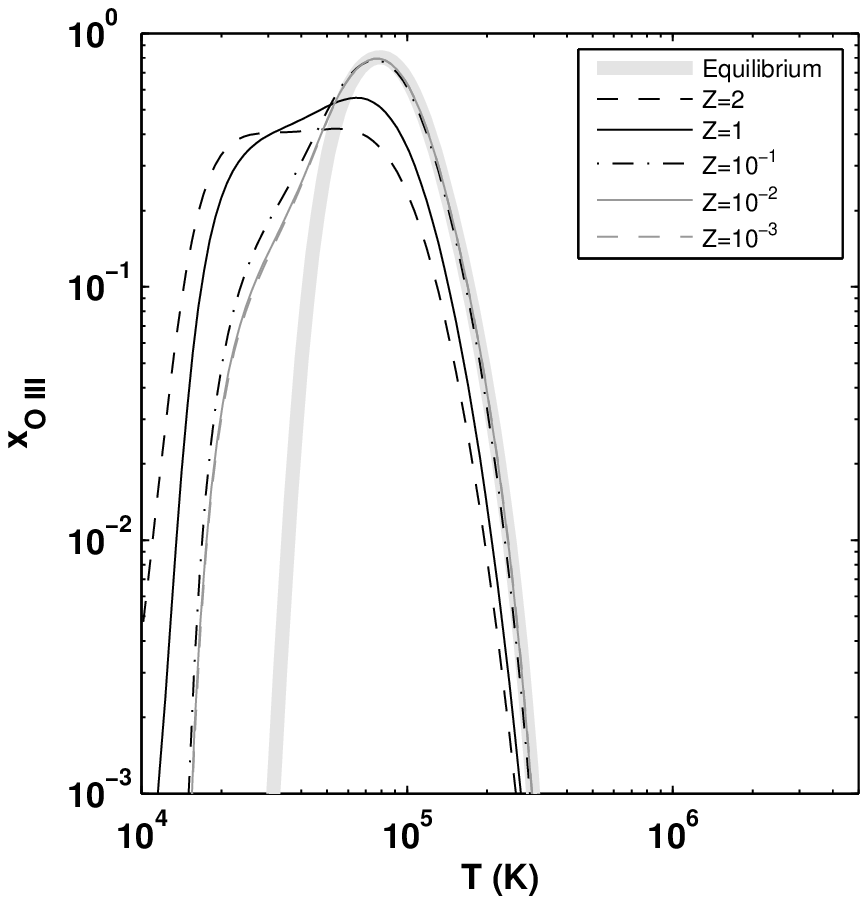}
\caption{O$^{5+}$ (left) and O$^{++}$ (right) ion fractions versus temperature
for metallicities $Z$ ranging from $10^{-3}$ to $2$,
for isochoric cooling. 
The CIE distributions are
shown by the thick grey lines.}
\label{OVI_Z}
\end{figure*}

\section{Cooling Efficiencies}
\label{cooling-res}
\begin{figure}
\epsscale{1.2}
\plotone{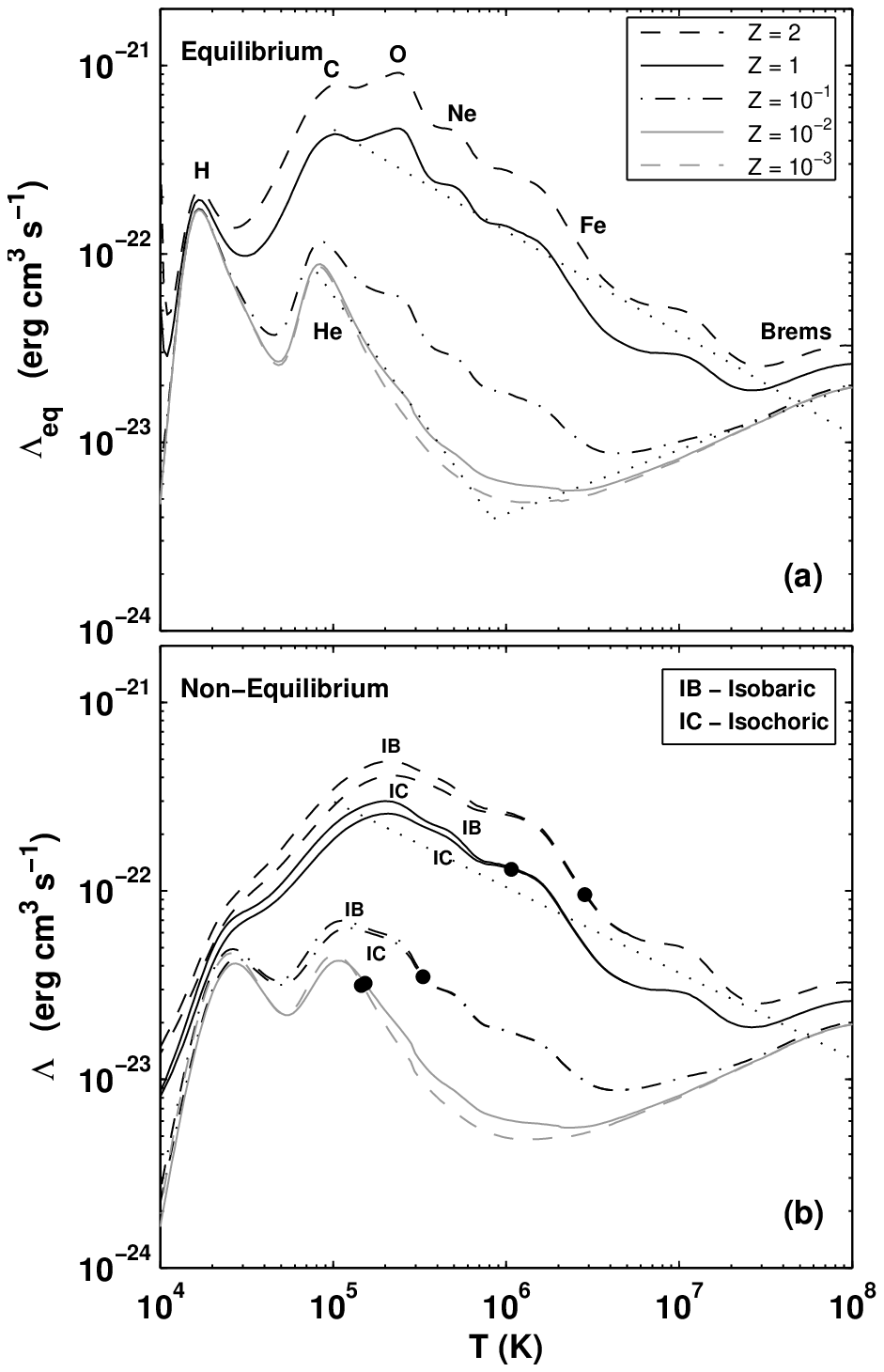}
\caption{Cooling efficiency (erg cm$^3$ s$^{-1}$) versus temperature
for $Z$ from $10^{-3}$ to $2$ times solar.
(a) Equilibrium cooling. The dominant cooling elements at various
temperatures are indicated near the curves.
The dotted lines are our power-law fits for $Z=1$ and $Z=10^{-3}$,
as given by equations \ref{pow1} and \ref{lowZfit}.
(b) Non-equilibrium cooling. 
For each $Z$, two curves are displayed: the lower
curve is for isochoric (IC) cooling, and the upper curve is 
isobaric (IB) cooling. 
The dotted line is our power-law  fit for $Z=1$ as given by equation
\ref{non-eq-fit}.
The filled circles indicate the temperatures where 
time-dependent cooling first deviate from the equilibrium cooling by 5\%.}
\label{cool-res-fig}
\end{figure}

We have carried out computations of the cooling efficiencies 
$\Lambda(T,x_i,Z)$ (erg cm$^3$ s$^{-1}$) 
assuming CIE at all temperatures, and for time-dependent
radiative cooling for isobaric and isochoric evolutions.
Our results are displayed in Figure \ref{cool-res-fig}, and
listed in Tables 13-15.
The upper panel of Figure \ref{cool-res-fig} shows our CIE cooling
functions, $\Lambda_{\rm eq}$, 
for $T$ between $10^4$ and $10^8$~K, and $Z$ from
$10^{-3}$ to $2$. The lower panel displays the
non-equilibrium cooling curves.

\setcounter{table}{12}
\begin{deluxetable}{llll}
\tablewidth{0pt}
\tablecaption{CIE Cooling Efficiencies}
\tablehead{
\colhead{Temperature} & 
\colhead{$\Lambda(Z=10^{-3})$} & 
\colhead{$\Lambda(Z=10^{-2})$} & 
\colhead{\ldots} \\
\colhead{(K)} &
\colhead{(erg cm$^3$ s$^{-1}$)} &
\colhead{(erg cm$^3$ s$^{-1}$)} &
\colhead{\ldots}}
\startdata
$1.00\times10^4$ & $4.70\times10^{-24}$ & $4.98\times10^{-24}$ & \ldots \\
$1.05\times10^4$ & $7.62\times10^{-24}$ & $7.79\times10^{-24}$ & \ldots \\
$1.10\times10^4$ & $1.23\times10^{-23}$ & $1.25\times10^{-23}$ & \ldots \\
$1.15\times10^4$ & $1.97\times10^{-23}$ & $1.98\times10^{-23}$ & \ldots \\
$1.20\times10^4$ & $3.08\times10^{-23}$ & $3.09\times10^{-23}$ & \ldots \\
\enddata
\tablecomments{
The complete version of this table is in the electronic edition of
the Journal.  The printed edition contains only a sample. }
\end{deluxetable}

In computing the CIE cooling efficiencies we use the 
equilibrium ion fractions $x_i^{\rm eq}(T)$ computed in \S 4
as inputs to the cooling functions.
Our CIE results reproduce the standard ``cosmic cooling 
curves'' presented in many
previous papers in the literature (see \S 1). 
Small
differences compared to previous calculations are expected due
to updates in the input atomic data 
and in the assumed gas phase abundances of the heavy elements.

For $Z\gtrsim0.1$,
the radiative energy losses between $10^5$ and $\sim 10^7$~K
are dominated by electron impact
excitations of resonance line transitions of metal ions.
Above $\sim 10^7$~K, metal line cooling becomes less effective, 
and bremsstrahlung radiation 
dominates.
The low-temperature peak at 
$2\times 10^4$~K is mainly due to hydrogen Ly$\alpha$ cooling,
a process that becomes less effective at higher temperatures 
where the neutral hydrogen fraction becomes small. 
Recombinations and forbidden-line transitions are minor
contributors to the gas cooling.
The familiar peaks and features in the cooling curves 
are due to specific metal line
coolants that dominate at different temperatures.
The Ly$\alpha$ maximum is followed by peaks at $1.0\times 10^5$,
$3.0\times 10^5$, $5.0\times 10^5$, and $1.5\times 10^6$,
due respectively to
resonance line transitions of carbon, oxygen, neon, and iron
ions.
For $Z=1$ we find a maximum equilibrium cooling efficiency of 
$4.6\times10^{-22}$~erg~cm$^3$~s$^{-1}$ at $T=2.3\times10^5$~K.

For $Z\lesssim 0.01$, metal cooling becomes negligible, 
and the energy losses are dominated
by hydrogen and helium only. H, He and He$^+$ line emissions
dominate at low $T$, and electron bremsstrahlung due to 
scattering with H$^+$ and He$^{++}$ at high $T$.
The cooling peak at $10^5$~K that appears when $Z$ becomes small, is
due to He$^+$ Ly$\alpha$. 

Between $10^5$ and $10^8$~K, our $Z=1$ cooling 
function is well fit
by the power-law expression (cf. Kahn et al.~1976),
\begin{equation}\label{pow1}
\Lambda_{\rm eq}^{Z=1} = 2.3\times10^{-19}~T^{-0.54}
\ \ \ \ {\rm erg \ cm^3 \ s^{-1} \ .}
\end{equation} 
This approximation is accurate to within a factor of $2.4$
over this temperature range ($10^5<T<10^8$~K), 
and to within a factor of 1.7 for $10^5<T<6.3\times10^7$~K. 
For $Z\gtrsim 1$, and between 10$^5$ and $\sim 10^7$~K,
$\Lambda$ is linearly proportional to $Z$. 
For this range of parameters it follows from equations~(\ref{pow1})
and (\ref{tcool}) that the cooling time 
$t_c\simeq 0.22~ T_6^{1.54}/(n\,Z)$ Myr
for isochoric cooling, and a factor 1.6 times longer for isobaric cooling.

For $Z\lesssim0.01$,  
the cooling efficiency is independent of $Z$.
In this low-Z limit, the two-piece power law
\begin{equation}
\Lambda_{\rm eq}^{Z=10^{-3}} = \left\{ \begin{array}{lcl}
   1.7\times10^{-16}~T^{-1.29}&,& 8\times10^4<T<8\times10^5~{\rm K}\\
   3.8\times10^{-26}~T^{~0.34}&,& 8\times10^5<T<10^8~{\rm K}
                  \end{array}\right.
\label{lowZfit}
\end{equation}
(in units of erg cm$^3$ s$^{-1}$)
reproduces the CIE cooling efficiency between $8\times 10^4$ and $10^8$~K
to within a factor of 1.35.

Figure \ref{cool-res-fig}b displays our results for the non-equilibrium 
cooling efficiencies. In these computations we use the non-equilibrium
ion fractions, $x_i(T)$, obtained by solving the coupled
equations (1) and (5), as presented in \S 4,
to compute $\Lambda(T,x_i,Z)$.
For each $Z$ we assume sufficiently large initial
temperatures such that ionization equilibrium is
established before significant cooling begins. Departures
from equilibrium then occur as the temperature drops and
the cooling times become short. For each $Z$ the filled circles
in Figure \ref{cool-res-fig}b
indicate the points where the cooling curves begin to
deviate by more than 5\% from CIE cooling. This is also where the isobaric
and isochoric cooling curves (indicated for each $Z$ by
the labels ``IB'' and ``IC'') bifurcate. 
For $Z=1$, departures from equilibrium set in for $T\lesssim10^6$~K.
The deviations
from CIE cooling begin at higher temperatures for higher $Z$,
and lower temperatures for lower $Z$.
For temperatures above the ``departure points'' the curves
in the upper and lower panels of Figure \ref{cool-res-fig}
are identical.

There are several important differences between the non-equilibrium
and CIE cooling curves. First, the distinct peaks and features that
appear for CIE are smeared out 
in the non-equilibrium cooling curves. 
This is due to the broader ion distributions, $x_i(T)$, 
that occur for non-equilibrium cooling. Individual ions
then contribute to the cooling over a larger temperature range.
For example, for non-equilibrium cooling at high $Z$
ions such as O$^{++}$ and Ne$^{++}$ persist below $10^5$~K, 
where they contribute significantly to the cooling in addition to Ly$\alpha$.
The Ly$\alpha$ cooling peak is itself broadened because a high
electron fraction is maintained down to low temperatures.

Second, 
the non-equilibrium cooling efficiencies are suppressed, by factors
of 2 to 4, compared to CIE cooling. 
For example, for $Z=1$ the maximum non-equilibrium cooling efficiency,
occurring at $T=2.1\times10^5$~K, is reduced to 
$2.6\times10^{-22}$~erg~cm$^3$~s$^{-1}$,
a factor of 1.8 less than the maximal CIE efficiency.
The suppression occurs because
the gas remains ``over-ionized'' as it cools, and the
densities of the specific
coolants that are most effective at each temperature are reduced. The
more highly ionized species that remain present at each temperature
generally have more energetic resonance line
transitions, and these are less efficiently excited
by the thermal electrons (McCray 1987).
For example, for non-equilibrium cooling 
in the low-$Z$ limit the absolute and relative heights
of the hydrogen and helium Ly$\alpha$ peaks are reduced
compared to CIE cooling. This is because the H$^0$ and He$^+$
fractions remain smaller at any temperature compared to CIE.

Third, because
isochoric cooling is faster (see equation [8])
with correspondingly greater recombination lags, the suppressions in the
cooling efficiencies are larger, and 
the isochoric curves fall below the isobaric curves.

The differences we find between CIE
and non-equilibrium cooling, are in good agreement with the results
found by Sutherland \& Dopita (1993; see their Fig.~13) and
Schmutzler \& Tscharnuter (1993; see their Fig.~2.2).
For $Z=1$ the power-law expression
\begin{equation}\label{non-eq-fit}
\Lambda_{\rm non-eq}^{Z=1}~=~5.6\times10^{-20}~T^{-0.46}
\ \ \ {\rm erg \ cm^3 \ s^{-1}} 
\end{equation}
provides a good fit to the non-equilibrium cooling efficiency
between $10^5$ and $10^8$~K, and is accurate to within a factor of 2.2.
For $10^5<T<5.5\times10^7$~K, it is accurate to within a factor of 1.5.
For low-$Z$, departures from CIE cooling
occur only below $2\times 10^5$~K, so that the fit given by
equation (\ref{lowZfit}) remains valid for radiatively cooling gas at
low metallicity, for $T>2\times 10^5$~K.
 
The suppression of the cooling efficiencies compared to CIE
implies longer cooling times. The actual recombination lag 
is therefore smaller than
would occur if the gas were able to cool at the faster CIE rates.
The departures from ionization equilibrium are thus stabilized
by the reduction in the cooling rates.

\section{Diagnostics}
\label{diag}
For a uniformly cooling gas cloud, 
the line-of-sight column densities of 
ions $i$ of element $m$ may be expressed simply as
$N_i^m=A_mx_i(T)N_{\rm H}$, where $N_{\rm H}$ is the
column density of hydrogen nuclei,
$A_m$ is the abundance of element $m$ (relative to H), and $x_i(T)$
are the temperature-dependent ion fractions,
as computed in \S\ref{ionization-res}.  
Specific column density ratios 
\begin{equation}\label{colrat}
\frac{N^m_i}{N^n_j} = \frac{A_m}{A_n}\frac{x_i(T)}{x_j(T)}
\end{equation}
may therefore be used as diagnostic probes of the ionization state,
temperature, and also metallicity in radiatively cooling gas.
For CIE, the ion fractions are independent of the
metallicity $Z$, and the column density ratios depend only on
the gas temperature (and the relative elemental abundances $A_m/A_n$).
However, for non-equilibrium cooling the ion fractions
also depend on $Z$, and hence so do the column density ratios.

\begin{figure*}
\epsscale{1.2}
\plotone{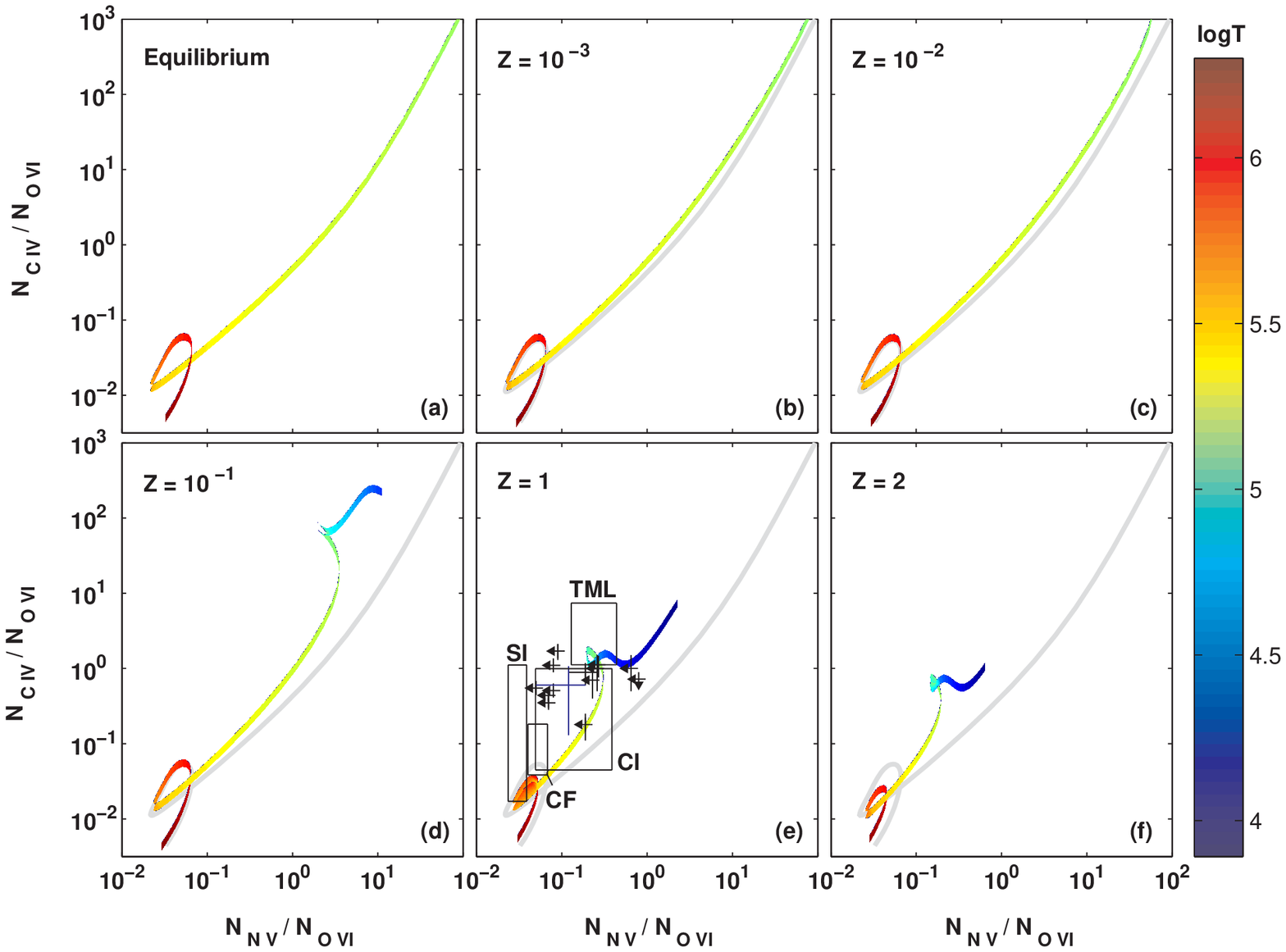}
\caption{Column density ratios
$N_{\rm C~IV} / N_{\rm O~VI}$ versus $N_{\rm N~V} / N_{\rm O~VI}$
for radiatively cooling gas. Gas temperature is indicated by color
along the trajectories, from hot (red) to cool (blue). Temperature
versus color legend is on the right. 
(a) displays the CIE trajectory, shown again as grey curves in the
other panels.
(b)-(f) are the trajectories non-equilibrium isochoric cooling
for different values of the metallicity $Z$.
The data points (Fox et al.~2005) show the ionic ratios observed in metal
absorbers toward HE 0226-4110 and PG 0953+414 (black), and in the Galactic
halo (blue). The boxes show model prediction, as demarcated by Fox et al., for
turbulent mixing layers (TMLs), conductive interfaces (CIs),
cooling flows (CFs) and shock ionization (SI).}
\label{dd}
\end{figure*}

Diagnostic diagrams for radiatively cooling gas for different
metallicities,
may be constructed using
the computational data we presented in \S 4 (Figs. 2-6 and
Tables 2-12).
As an example
\footnote{Diagnostic diagrams for any combination of ion-ratios
may be constructed automatically using our web tool at
http://wise-obs.tau.ac.il/$\sim$orlyg/cooling/.},
in Figure \ref{dd} we display 
``cooling trajectories'' for $N_{\rm C~IV} / N_{\rm O~VI}$ versus 
$N_{\rm N~V} / N_{\rm O~VI}$, for $T$ ranging from 
$5\times 10^6$ to $10^4$ K. 
The gas temperature is represented by color along the
curves, from hot (red) to cool (blue).  
Panel (a) shows the behavior
for CIE, for which the results are independent of $Z$.
Panels (b)-(f) are
for non-equilibrium isochoric cooling, for $Z$ from
$10^{-3}$ to $2$. 
The CIE trajectory is reproduced as the grey curves in panels (b)-(f).

For the  $Z$=1 model in Figure \ref{dd} we also display, as 
observational examples, the data points
and upper limits presented by Fox et al.~(2005) 
for the multiphased high-velocity cloud
absorbers towards the background sources 
HE 0226-4110 and PG 0953+414 (black crosses). The plot also includes the 
Galactic halo average (blue cross) 
$N_{\rm C~IV}/ N_{\rm O~VI}=0.6\pm0.47$ and 
$N_{\rm N~V} / N_{\rm O~VI}=0.12\pm0.07$
(Zsarg{\'o} et al.~2003, as quoted by Fox et al.). 
We also
show the regions demarcated by Fox et al.~for
the predictions for a variety of ionization mechanisms,
including turbulent mixing layers (TMLs; Slavin, Shull, \& Begelman 1993), 
conductive interfaces (CIs; Borkowski, Balbus, \& Fristrom 1990),
cooling flows (CFs; Edgar \& Chevalier 1986) 
and shock ionization (SI; Dopita \& Sutherland 1996).

\begin{figure*}
\epsscale{1.2}
\plotone{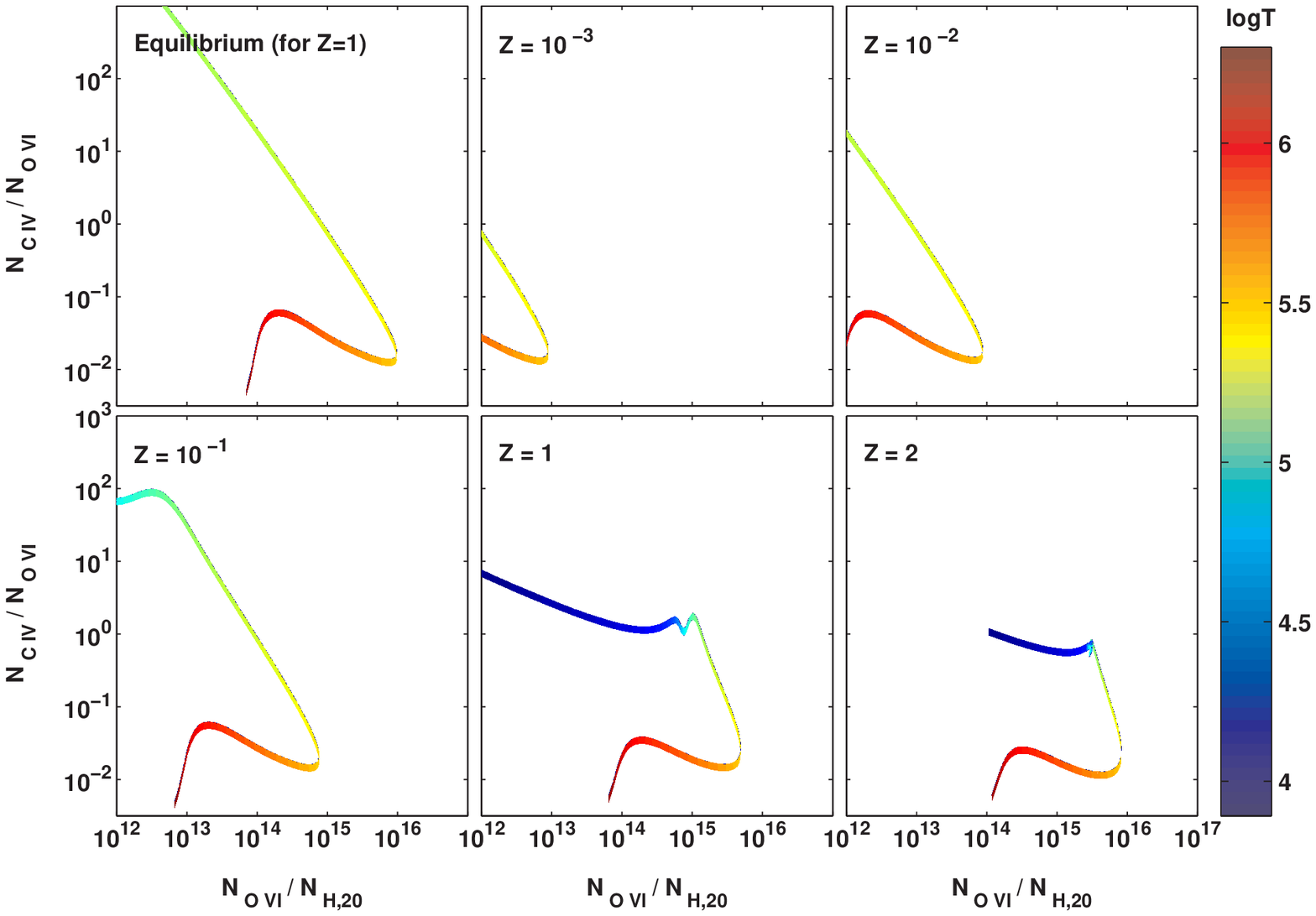}
\caption{Same as Figure \ref{dd} but for column density ratio 
$N_{\rm C~IV} / N_{\rm O~VI}$ versus $N_{\rm O~VI} / N_{{\rm H},20}$,
where $N_{{\rm H},20}$ is the hydrogen column density in units of 
$10^{20}$~cm$^{-2}$.}
\label{dd-abs}
\end{figure*}

To set the absolute scale of our predicted ion columns, 
in Figure \ref{dd-abs} we plot $N_{\rm C~IV} / N_{\rm O~VI}$ versus
$N_{\rm O~VI} / N_{{\rm H},20}$ for each $Z$, where $N_{{\rm H},20}$ is the
hydrogen column density in units of $10^{20}$~cm$^{-2}$.  
For the CIE calculation we assume $Z=1$. 
At any temperature, the \ion{O}{6} column scales linearly 
with the assumed $N_{\rm H}$, whereas the column density ratios
are independent of $N_{\rm H}$.

Of the three ions we are considering, O$^{5+}$ is the most highly ionized,
so that C~{\small IV}/O~{\small VI} and N~{\small V}/O~{\small VI} are
small at high temperature, and become large at low temperature.
Figures \ref{dd} and \ref{dd-abs} show again
that for time-dependent radiative cooling the gas
remains more highly ionized compared to CIE.
As $Z$ becomes large and the recombination lags grow, 
the ion ratios remain small even at low temperatures.
The cooling trajectories are therefore
confined to a narrower range of ion ratios
for non-equilibrium cooling.
Furthermore, in our example, it is evident that for a
given N~{\small V}/O~{\small VI} ratio, the associated 
C~{\small IV}/O~{\small VI} ratio is larger for non-equilibrium
cooling compared to CIE. 
For example, a cloud for which
$N_{\rm N~V}/N_{\rm O~VI} \simeq 3$ and
$N_{\rm C~IV}/N_{\rm O~VI} \simeq 30$ would be 
inconsistent with gas at CIE at any temperature, yet 
consistent with isochoric radiatively cooling gas at a
temperature near $\sim1.3\times 10^5$~K.
It is clear that assuming CIE
when interpreting such data may lead to 
significant inaccuracies.
For example, for an isochorically cooling cloud with $Z\simeq 1$, 
absorption line measurements yielding
$N_{\rm C~IV}/N_{\rm O~VI} \approx 1$ and 
$N_{\rm N~V}/N_{\rm O~VI} \approx 1$ would be interpreted as indicating
$T\simeq2\times10^5$~K assuming CIE, whereas in fact 
these ratios occur at a much lower temperature $\sim10^4$~K.

Interestingly, many (but not all) of the Fox et al.~(2005)
data points lie close to
our $Z=1$ cooling trajectory (see Figure \ref{dd}e). 
At lower metallicities the column density ratios
approach the CIE trajectory which lies well below the data points.
However, 
as indicated by Figure \ref{dd}e, even for $Z=1$ 
further constraints may be required to distinguish radiatively cooling gas 
from ionization occurring in
conductive interfaces or turbulent mixing layers.

\section{Summary}
\label{sum}
In this paper we present new computations of the equilibrium and
non-equilibrium cooling efficiencies and ionization states for
low-density radiatively cooling gas, containing
cosmic abundances of the elements H, He, C, N, O, Ne, Mg, Si, 
S, and Fe.  
In these calculations we assume pure radiative cooling, 
with no gas heating or photoionization by external sources. 
We present results for gas
temperatures, $T$, between 10$^4$ and 10$^8$~K. We assume that the
gas is dust free, and we consider
metallicities $Z$ ranging from 10$^{-3}$ to 2 times
the elemental abundances in the Sun. We carry out
our computations using up-to-date rate coefficients
for all of the atomic recombination and
ionization processes, and the energy loss mechanisms. 

For temperatures below $\sim 5\times 10^6$~K, where ion-electron recombination
lags significantly behind the cooling, we explicitly solve the
coupled time-dependent ionization and energy loss equations for the
cooling gas. For such gas we assume that
the cooling is from an initially hot equilibrium state.
The basic equations and our numerical method are presented in \S 2.
We calculate the non-equilibrium
cooling efficiencies for constant pressure 
(isobaric) and constant density (isochoric) evolutions. Departures
from collisional ionization equilibrium (CIE)
are slightly smaller for isobaric cooling.

In \S 3 we consider the conditions for isochoric versus
isobaric cooling. We compute the critical column densities 
and temperatures at which the cooling time becomes
short compared to the dynamical time, and  
the transition from isobaric to isochoric evolution occurs.
These results are displayed in Figure \ref{choose-case}, and
are based on the cooling efficiencies we present in \S 5.

Because we exclude photoionization by external radiation,
both the equilibrium and time-dependent behavior is independent
of the assumed density or pressure. The primary parameter
for radiative cooling is the metallicity.
Departures from equilibrium 
are largest for high $Z$ where the ratios of the
cooling and recombinations times are smallest.
At very low metallicity ($Z\lesssim 10^{-2}$) the cooling
rates, and hence also the recombination lags,
are independent of the metallicity. In this
limit any departures from CIE
approach a universal ``primordial'' form.
The results of our equilibrium ionization calculations
are displayed in Figure 2 in \S 4. This figure
also shows the ionization states for non-equilibrium
isobaric and isochoric cooling, for $Z=1$. 
Results for other metallicities are presented
in on-line tables and figures, as summarized in
Table 16. Our computational data set can also be
found at our website http://wise-obs.tau.ac.il/$\sim$orlyg/cooling/.

\setcounter{table}{15}
\begin{deluxetable}{llcc}
\tablewidth{0pt}
\tablecaption{Ionization and Cooling Data}
\tablehead{
\colhead{Data} & \colhead{} &
\colhead{Table} & 
\colhead{Figure}}
\startdata
\multicolumn{2}{l}{Ion Fractions:} &&\\
~~~CIE                   & & $2$  & $2$\\
~~~$Z=1$, &Isochoric       & $3$  & $2$  \\
~~~$Z=1$, &Isobaric        & $4$  & $2$  \\
~~~$Z=10^{-3}$, &Isochoric & $5$  & $3$  \\
~~~$Z=10^{-3}$, &Isobaric  & $6$  & $3$  \\
~~~$Z=10^{-2}$, &Isochoric & $7$  & $4$  \\
~~~$Z=10^{-2}$, &Isobaric  & $8$  & $4$  \\
~~~$Z=10^{-1}$, &Isochoric & $9$  & $5$  \\
~~~$Z=10^{-1}$, &Isobaric  & $10$ & $5$  \\
~~~$Z=2$, &Isochoric       & $11$ & $6$  \\
~~~$Z=2$, &Isobaric        & $12$ & $6$  \\
\multicolumn{2}{l}{Cooling Efficiencies:} &&\\
~~~CIE                    && $13$ & $8$  \\
~~~Isochoric       && $14$ & $8$  \\
~~~Isobaric        && $15$ & $8$  \\
\enddata
\label{tables}
\end{deluxetable}

Figure \ref{cool-res-fig} in \S 5 displays our results for the
equilibrium and non-equilibrium cooling efficiencies
for $Z$ between 10$^{-3}$ and $2$.
We provide simple, or two-piece, power-law fits 
(equations [14], [15], and [16]) for the
equilibrium and non-equilibrium cooling efficiencies at
high and low $Z$. Non-equilibrium cooling is suppressed,
by factors of 2 to 4, relative to CIE, because of the
generally higher ionization state of the rapidly cooling gas.
The familiar peaks in the CIE cooling curve are smeared
out for non-equilibrium cooling, because individual ions
persist for a broader range of temperatures. Overall, 
our results are in good agreement with previous
such calculations. Detailed difference are mainly due to
differences in the input atomic data, and assumed abundances
of the heavy elements. We make some explicit comparisons
with previous computations in \S 4 and \S 5.

Ion ratios are useful as diagnostic probes.
In \S 6 we discuss one example,
N~{\small V}/O~{\small VI} versus C~{\small IV}/O{\small VI},
and show how this ratio evolves in
radiatively cooling gas, and how it can be used as
a probe of metallicity for realistic non-equilibrium conditions.

In our computations we assume 
that the cooling gas is optically thin.
In the Appendix,
we provide numerical estimates for the maximal cloud column densities for
which this assumption remains valid.
We also investigate how reabsorption of hydrogen and
helium recombination radiation in optically thick clouds
alters the cooling rates and associated ionization states
in non-equilibrium cooling gas.
For high metallicity the effects of trapping
are small, but become more significant
($\sim$~factors of 2) for low $Z$ gas.

In a companion paper we will present computations of the metal ion
``cooling columns'' produced in steady flows of radiative
cooling gas, such as occur in post shock cooling layers,
including the effects of ``upstream'' cooling radiation on
the ``downstream'' ionization states. 
We will also consider how photoionization by external
background radiation fields
alters the ionization
and thermal evolution of radiatively cooling gas such as
we have considered here.

\section*{Acknowledgments}
We thank Gary Ferland for his invaluable assistance in our
non-standard use of Cloudy. We thank Hagai Netzer
for generously providing us with his up-to-date ION atomic data set.
We thank Chris McKee for many helpful discussions.
Our research is supported by the US-Israel Binational
Science Foundation (grant 2002317).

\appendix
\section{Reabsorption of Diffuse Radiation Fields}
\label{diffuse}
In the computations presented in \S 4 and \S 5 we have assumed that
the cooling gas is optically thin, and that  
reabsorption of line and continuum radiation emitted by the cooling gas
may be ignored.
In this Appendix we provide numerical estimates
for the maximal cloud column densities for which the
optically thin assumption is justified.
We then examine how trapping of
hydrogen and helium recombination radiation in optically
thick clouds affects the time-dependent 
cooling efficiencies and ionization states.

For sufficiently large cloud column densities, absorption of the 
internally generated ``diffuse radiation'' can alter the ionization
state in such a way as to enhance or reduce the net cooling rate.  
To estimate the critical column densities at which 
the cooling rates are altered, we use
Cloudy to compute the local CIE cooling rates as functions
of the total hydrogen column density, $N_{\rm H}$,
for one-dimensional constant temperature slabs.
We set the external radiation field to zero, so that the 
radiative transfer handled by Cloudy is for the 
internally generated diffuse fields only. For each temperature, $T$, and 
assumed metallicity $Z$, we identify the critical column density,
$N_{\rm Hcrit}(T,Z)$, at which the local cooling efficiency 
at the cloud center first deviates
by $50\%$ from the optically thin cooling rate at the cloud edge.

\begin{figure}
\epsscale{0.5}
\plotone{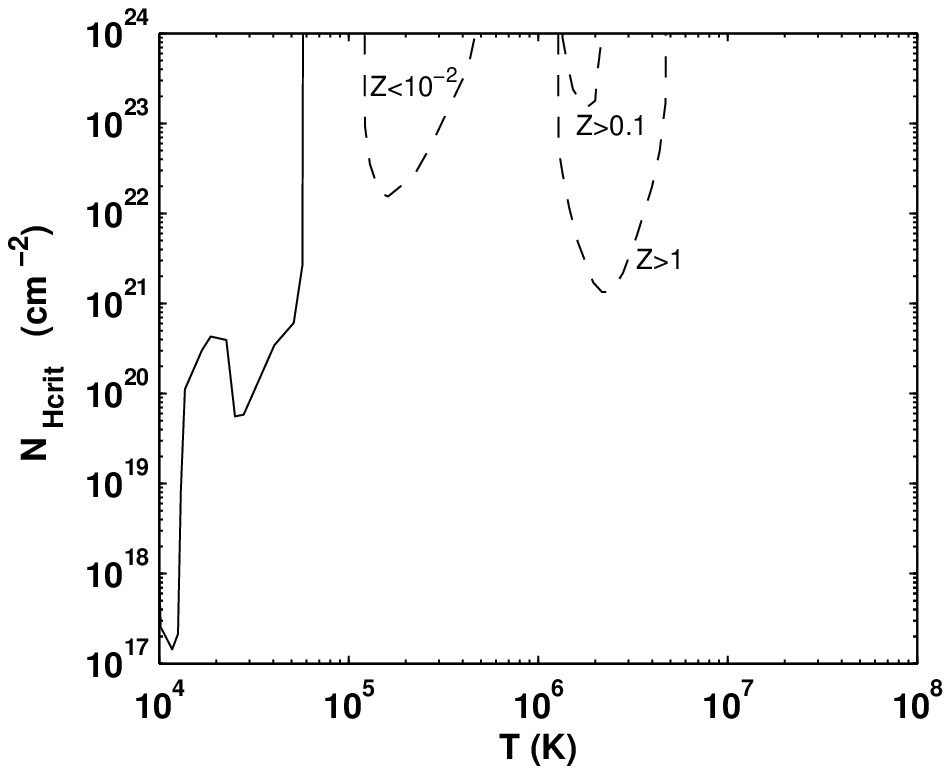}
\caption{$N_{\rm Hcrit}$ versus temperature. $N_{\rm Hcrit}$ is
the column density at which the cooling deviates by  
$50\%$ from the optically thin value (see text). 
The solid curve is for all metallicities $10^{-3}<Z<2$. 
The dashed curves are for the displayed $Z$ range.
For temperatures where no curve is displayed,  
$N_{\rm Hcrit}> 10^{24}$~cm$^{-2}$. 
}
\label{Ncrit}
\end{figure}

The critical column densities defined in this way are sensitive to
the gas temperature and associated ionization states. Here
we are not interested in the detailed fluctuations of $N_{\rm Hcrit}$
with gas temperature, but rather with the broad trends. 
In Figure \ref{Ncrit}, we display a smoothed representation of
$N_{\rm Hcrit}$ versus $T$. The 
solid curve, appearing for $T\lesssim 6\times 10^4$~K
is for the entire metallicity range $10^{-3}<Z<2$. 
For higher temperatures, the behavior is sensitive to $Z$,
and we display the critical columns by dashed lines
for selected metallicity ranges. 
For temperatures where
no curves appear in Figure \ref{Ncrit} the critical columns exceed
$10^{24}$~cm$^{-2}$.

Three distinct regimes appear in Figure \ref{Ncrit}. 
For $10^4\lesssim T \lesssim6\times10^4$~K, 
a transition from ``case A'' to ``case B'' hydrogen recombination 
occurs as the cloud becomes optically thick. Photoionization
of hydrogen by H$^+$ and He$^+$ recombination radiation then
alters the dominating Ly$\alpha$ cooling rate.
The Ly$\alpha$ cooling rate is proportional to the product 
$x_{\rm e}(1-x_{\rm e})$,
where $x_{\rm e}$ is the electron fraction.
For $T\lesssim 1.5\times 10^4$~K, the neutral fraction
$(1-x_{\rm e})\approx 1$,  
whereas $x_{\rm e}$ increases in the transition to case B.
This leads to {\it increased}
Ly$\alpha$ cooling. At these low temperatures we find that 
$N_{\rm Hcrit} \sim10^{17}$~cm$^{-2}$.
At higher temperatures, $T\gtrsim 2.5\times 10^4$~K,
the hydrogen is largely ionized and $x_{\rm e}\approx 1$ so 
the transition to case B mainly reduces the neutral fraction $(1-x_{\rm e})$.
This leads to {\it decreased} Ly$\alpha$ cooling. 
For $T= 2.5\times 10^4$~K, we find
$N_{\rm Hcrit}\approx 5\times 10^{19}$ cm$^{-2}$.
Between these two temperatures, $x_{\rm e}(1-x_{\rm e})$ remains
approximately constant in the transition to case B,
and a local maximum of $N_{\rm Hcrit}=4\times 10^{20}$~cm$^{-2}$ 
appears at $T\approx2\times10^4$~K.
Above $T\sim 6\times 10^4$~K
$N_{\rm Hcrit}$ rises sharply because 
the hydrogen becomes so highly ionized
that Ly$\alpha$ cooling is no longer the dominant coolant.  

Similarly, absorption of He$^{++}$ recombination radiation
by He$^+$ ions alters the He$^+$ Ly$\alpha$ cooling rate
at temperatures near 10$^5$~K where helium makes
the transition from singly to doubly ionized form.
However, this effect is only important
for low metallicity ($Z < 10^{-2}$) clouds, where
He$^+$ Ly$\alpha$ is a significant coolant.
For such clouds, $N_{\rm Hcrit}=1.5\times10^{22}$~cm$^{-2}$ at 
$T=1.6\times10^5$~K.

Finally, for $10^6 \lesssim T \lesssim 5\times10^6$~K, we find that
$N_{\rm Hcrit}\sim 10^{21}$~cm$^{-2}$ in 
high metallicity ($Z~\gtrsim~1$) clouds. At these temperatures,  
photoionization by numerous energetic metal emission lines 
ionize O$^{7+}$, 
and shift the peak of the iron ion distribution from Fe$^{11+}$-Fe$^{14+}$
to Fe$^{13+}$-Fe$^{16+}$. This reduces the cooling rate,
because O$^{7+}$, Fe$^{11+}$, and Fe$^{12+}$ are dominant coolants 
at these temperatures.
For lower $Z$, bremsstrahlung cooling plays a more dominant role,
and photoionization of the oxygen and iron ions is less important,
so $N_{\rm Hcrit}$ is much larger.

Our computations for $N_{\rm Hcrit}$ in Figure \ref{Ncrit} are
for CIE conditions. 
We now examine how trapping of hydrogen and helium
recombination radiation alters the cooling efficiencies and non-equilibrium 
ionization states for time-dependent cooling. 
We consider a series of isochoric model clouds with 
total column densities ranging up to 
$N_{\rm H} = 10^{24}$~cm$^{-2}$. 
We assume that the H$^+$, He$^+$, and He$^{++}$
recombination photons are reabsorbed on-the-spot 
when the photoionization optical depths
at the H$^0$, He$^0$ and He$^+$ ionization thresholds
exceed unity respectively.  The critical temperatures below which the clouds
become optically thick 
are higher for clouds with larger total column densities, since 
lower neutral hydrogen and helium fractions are required.
When the clouds become optically thick we
switch from case A to case B recombination 
in computing the evolution of the hydrogen and helium ionization 
states\footnote{For He$^+$ recombination radiation, we
assume that an appropriate fraction $y$ of photons emitted in
 recombinations to the
He ground state are absorbed by neutral H, and a fraction $(1-y)$ are
absorbed by neutral He. We adopt the low-density limit $p=0.96$ for
the fraction of helium recombinations
to excited states that are absorbed on-the-spot by neutral H
(see Osterbrock 1989).}.
 
\begin{figure}
\epsscale{0.9}
\plotone{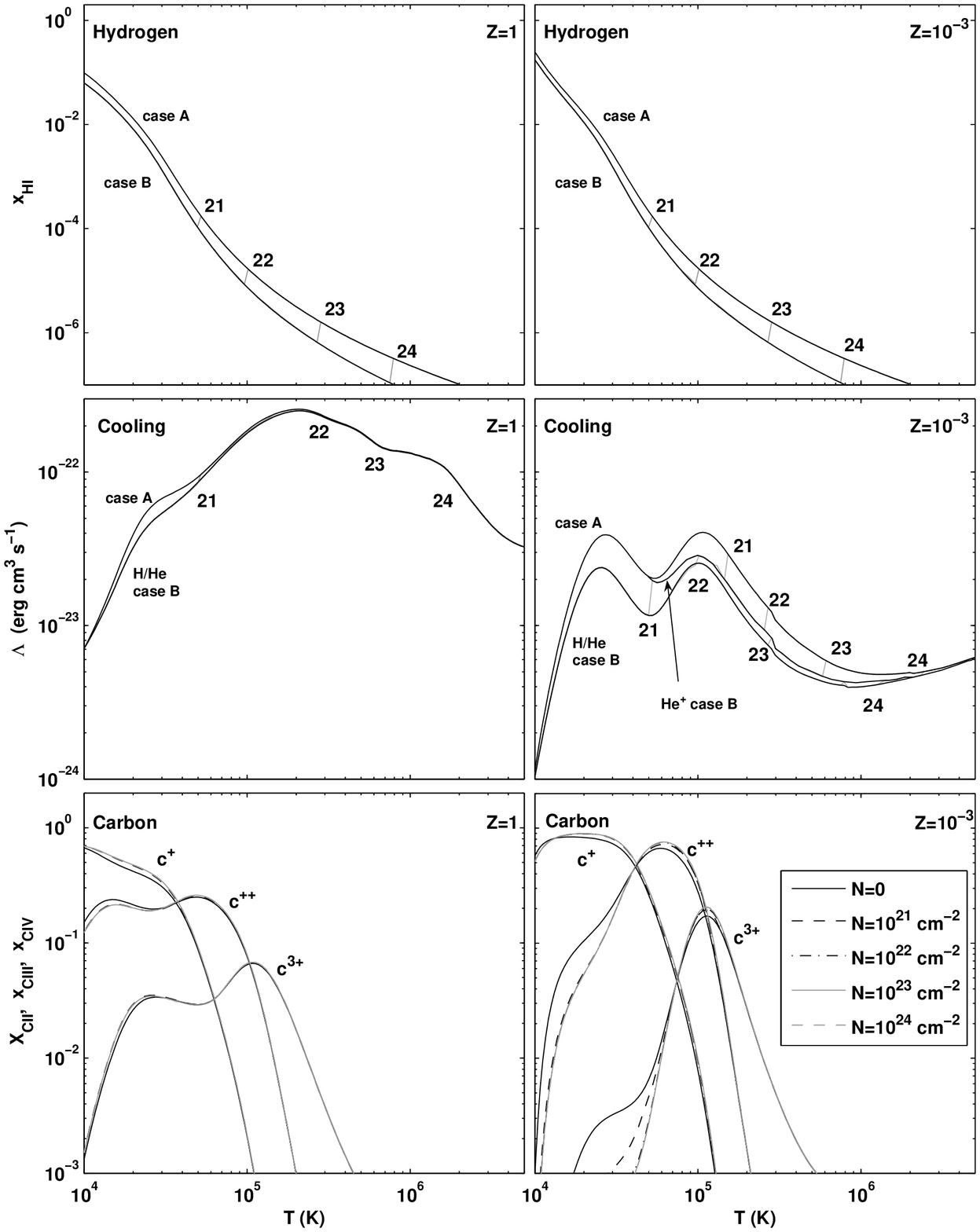}
\caption{Ionization and cooling in optically thin and thick clouds.
Left hand panels are for $Z=1$. Right hand panels are for $Z=10^{-3}$.
Upper row shows the neutral hydrogen fractions versus temperature
for optically thin ``case A'', and optically thick ``case B'' (see text).
The vertical lines mark the temperatures at which the gas
becomes optically thick for different total cloud column densities, 
as indicated by the adjacent labels showing $log(N_{\rm H})$.
Middle row displays the optically thin and thick cooling efficiencies.
For $Z=10^{-3}$ the intermediate curve is for a cloud that is
optically thick only to He$^{++}$ recombination radiation (see text).
Lower row displays the C$^+$, C$^{++}$, and C$^{3+}$ ion fractions
for optically thin and thick clouds.}
\label{caseB-fig}
\end{figure}

In Figure \ref{caseB-fig} we plot the neutral hydrogen fractions,
$x_{\rm H}(T)$,
and cooling efficiencies, $\Lambda(T)$,
for case A and case B recombination. 
We display results for $Z=1$ (left-hand panels), 
and for $Z=10^{-3}$ (right-hand panels).
To illustrate the effects 
on the metal ion distributions we also plot
the C$^+$, C$^{++}$, and C$^{3+}$ fractions.

The neutral hydrogen fractions are, of course, reduced
in optically thick clouds.
The vertical lines connecting the case A
and case B curves indicate the temperatures at which, for
a given total cloud column $N_{\rm H}$,
the cloud becomes optically thick
to hydrogen recombination radiation.
For example, for $N_{\rm H}=10^{21}$ cm$^{-2}$,
this occurs at $T=5\times10^4$~K.

For $Z=1$ the differences between case A and case B cooling are
small, and are only apparent at low temperatures where Ly$\alpha$
cooling dominates. For non-equilibrium cooling
the electron fraction remains large down to 
$10^4$~K. Therefore, down to this temperature Ly$\alpha$ cooling is reduced 
by the reduction in the neutral H fraction.
(This is opposed to the situation in CIE where, as discussed above,
 for $10^4$~K gas
the Ly$\alpha$ cooling is {\it enhanced} by
an increased electron density in optically thick clouds.)
For $Z=1$, the slightly altered cooling efficiencies in 
optically thick clouds lead to only very small changes 
in the non-equilibrium metal ion abundances, as is 
illustrated for the carbon ions in Figure \ref{caseB-fig}.
Our conclusions for $Z=1$ are consistent with the results
of Kafatos (1973) and Shapiro \& Moore (1976), who considered
the transition from optically thin to thick 
(for hydrogen recombination radiation only) at 
a single temperature $T=3.5\times 10^4$~K
(corresponding to $N_{\rm H}\approx1.3\times10^{20}$~cm$^{-2}$).

For $Z=10^{-3}$, 
hydrogen and helium are already
the dominant emission line coolants below $\sim 10^6$~K,
and the transition to optically thick conditions has a larger effect.
The cooling efficiencies are reduced by up to a factor $\sim 2$.
In Figure \ref{caseB-fig} we draw three
cooling curves for $Z=10^{-3}$. 
The upper curve is for optically 
thin case A cooling. The middle curve (labeled ``He$^{+}$ case B'')
is for clouds that are still optically thin to H$^+$ and He$^+$ 
recombination radiation, but are thick to He$^{++}$ recombination photons.
The lower curve is for case B clouds that are thick to the 
H$^+$ and He$^+$ recombination photons as well.
The vertical lines connecting the curves indicate the
transition temperatures for different cloud columns $N_{\rm H}$.
For example, a $10^{21}$~cm$^{-2}$ cloud 
shifts from case A to ``He$^{+}$ case B'' at $\sim2\times10^5$~K, and
then becomes thick to H$^+$ and He$^+$ recombination radiation
(``H/He case B'') at $\sim5\times10^4$~K.

For low-$Z$ gas the overall cooling rates are small, and hence
so are the departures from CIE. Therefore, even though the
cooling efficiencies are significantly altered by trapping of
the recombination radiation, the resulting effect on the
ion fractions is not very large. As shown for 
C$^+$, C$^{++}$, and C$^{3+}$ the effects are most significant at
low temperatures where the departures from CIE are largest.

\clearpage

\begin{thebibliography}{}

\bibitem[Aldrovandi \& Pequignot(1973)]{1973A&A....25..137A} Aldrovandi,
S.~M.~V., \& Pequignot, D.\ 1973, \aap, 25, 137

\bibitem[Allen \& Dupree(1969)]{1969ApJ...155...27A} Allen, J.~W., \& 
Dupree, A.~K.\ 1969, \apj, 155, 27 

\bibitem[Altun et al.(2004)]{2004A&A...420..775A} Altun, Z., Yumak, A.,
Badnell, N.~R., Colgan, J., \& Pindzola, M.~S.\ 2004, \aap, 420, 775

\bibitem[Altun et al.(2005)]{2005A&A...433..395A} Altun, Z., Yumak, A.,
Badnell, N.~R., Colgan, J., \& Pindzola, M.~S.\ 2005, \aap, 433, 395

\bibitem[Altun et al.(2006)]{2006A&A...447.1165A} Altun, Z., Yumak, A.,
Badnell, N.~R., Loch, S.~D., \& Pindzola, M.~S.\ 2006, \aap, 447, 1165

\bibitem[Anders \& Grevesse(1989)]{1989GeCoA..53..197A} Anders, E., \& 
Grevesse, N.\ 1989, \gca, 53, 197 

\bibitem[Antia \& Basu(2005)]{2005ApJ...620L.129A} Antia, H.~M., \& Basu, 
S.\ 2005, \apjl, 620, L129 

\bibitem[Arnaud \& Raymond(1992)]{1992ApJ...398..394A} Arnaud, M., \&
Raymond, J.\ 1992, \apj, 398, 394

\bibitem[Arnaud \& Rothenflug(1985)]{1985A&AS...60..425A} Arnaud, M., \& 
Rothenflug, R.\ 1985, \aaps, 60, 425 

\bibitem[Asplund et al.(2005)]{2005ASPC..336...25A} Asplund, M., Grevesse, 
N., \& Sauval, A.~J.\ 2005a, ASP Conf.~Ser.~336: Cosmic Abundances as 
Records of Stellar Evolution and Nucleosynthesis, 336, 25 

\bibitem[Asplund et al.(2005)]{2005astro.ph.10377A}Asplund, M., Grevesse, N., 
Gudel, M., \& Sauval, A.J.\ 2005b, astro-ph/0510377

\bibitem[Ayres et al.(2006)]{2006ApJS..165..618A} Ayres, T.~R., Plymate, 
C., \& Keller, C.~U.\ 2006, \apjs, 165, 618 

\bibitem[Badnell et al.(2003)]{2003A&A...406.1151B} Badnell, N.~R., et al.\ 2003, \aap, 406, 1151

\bibitem[Badnell(2006)]{2006A&A...447..389B} Badnell, N.~R.\ 2006, \aap,
447, 389

\bibitem[Ballantyne et al.(2000)]{2000ApJ...536..773B} Ballantyne, D.~R., 
Ferland, G.~J., \& Martin, P.~G.\ 2000, \apj, 536, 773 

\bibitem[Bahcall et al.(2005)]{2005ApJ...631.1281B} Bahcall, J.~N., Basu, 
S., \& Serenelli, A.~M.\ 2005, \apj, 631, 1281 

\bibitem[Benjamin et al.(2001)]{2001ApJ...554L.225B} Benjamin, R.~A., 
Benson, B.~A., \& Cox, D.~P.\ 2001, \apjl, 554, L225 

\bibitem[Boehringer \& Hensler(1989)]{1989A&A...215..147B} Boehringer, H., 
\& Hensler, G.\ 1989, \aap, 215, 147 

\bibitem[Borkowski et al.(1990)]{1990ApJ...355..501B} Borkowski, K.~J., 
Balbus, S.~A., \& Fristrom, C.~C.\ 1990, \apj, 355, 501 

\bibitem[Bottcher et al.(1970)]{01}Bottcher, C., McCray, R.A., Jura, M., \& Dalgarno, A.\
1970, Ap.Lett., 6, 237

\bibitem[Breitschwerdt \& Schmutzler(1994)]{1994Natur.371..774B} 
Breitschwerdt, D., \& Schmutzler, T.\ 1994, \nat, 371, 774 

\bibitem[Breitschwerdt \& Schmutzler(1999)]{02}Breitschwerdt, D., \& Schmutzler, T.\ 1999, \aa, 347, 650

\bibitem[Cen \& Ostriker(1999)]{1999ApJ...514....1C} Cen, R., \& Ostriker, 
J.~P.\ 1999, \apj, 514, 1

\bibitem[Clarke et al.(1998)]{03}Clarke, N.J., et al.\ 1998, J. Phys. B: At. Mol. Opt. Phys. 33, 533

\bibitem[Colgan et al.(2003)]{2003A&A...412..597C} Colgan, J., Pindzola,
M.~S., Whiteford, A.~D., \& Badnell, N.~R.\ 2003, \aap, 412, 597

\bibitem[Colgan et al.(2004)]{2004A&A...417.1183C} Colgan, J., Pindzola,
M.~S., \& Badnell, N.~R.\ 2004, \aap, 417, 1183

\bibitem[Colgan et al.(2005)]{2005A&A...429..369C} Colgan, J., Pindzola,
M.~S., \& Badnell, N.~R.\ 2005, \aap, 429, 369

\bibitem[Collins et al.(2005)]{2005ApJ...623..196C} Collins, J.~A., Shull, 
J.~M., \& Giroux, M.~L.\ 2005, \apj, 623, 196 

\bibitem[Cox \& Tucker(1969)]{1969ApJ...157.1157C} Cox, D.~P., \& Tucker, 
W.~H.\ 1969, \apj, 157, 1157 

\bibitem[Dalgarno \& McCray(1972)]{1972ARA&A..10..375D} Dalgarno, A., \& 
McCray, R.~A.\ 1972, \araa, 10, 375 

\bibitem[Dav{\'e} et al.(2001)]{2001ApJ...552..473D} Dav{\'e}, R., et al.\ 
2001, \apj, 552, 473 

\bibitem[Draine(1981)]{1981ApJ...245..880D} Draine, B.~T.\ 1981, \apj, 245, 
880 

\bibitem[Draine \& McKee(1993)]{04}Draine, B.T., \& McKee, C.F.\ 1993, \araa, 31, 373

\bibitem[Dopita \& Sutherland(1996)]{1996ApJS..102..161D} Dopita, M.~A., \& 
Sutherland, R.~S.\ 1996, \apjs, 102, 161 

\bibitem[Drake \& Testa(2005)]{2005Natur.436..525D} Drake, J.~J., \& Testa, 
P.\ 2005, \nat, 436, 525 

\bibitem[Edgar \& Chevalier(1986)]{1986ApJ...310L..27E} Edgar, R.~J., \& 
Chevalier, R.~A.\ 1986, \apjl, 310, L27 

\bibitem[Fang et al.(2006)]{2006ApJ...644..174F} Fang, T., Mckee, C.~F., 
Canizares, C.~R., \& Wolfire, M.\ 2006, \apj, 644, 174 

\bibitem[Ferland et al.(1997)]{1997ApJ...481L.115F} Ferland, G.~J., 
Korista, K.~T., Verner, D.~A., \& Dalgarno, A.\ 1997, \apjl, 481, L115 

\bibitem[Ferland et al.(1998)]{1998PASP..110..761F} Ferland, G.~J., 
Korista, K.~T., Verner, D.~A., Ferguson, J.~W., Kingdon, J.~B., \& Verner, 
E.~M.\ 1998, \pasp, 110, 761

\bibitem[Fox et al.(2004)]{2004ApJ...602..738F} Fox, A.~J., Savage, B.~D., 
Wakker, B.~P., Richter, P., Sembach, K.~R., \& Tripp, T.~M.\ 2004, \apj, 
602, 738 

\bibitem[Fox et al.(2005)]{2005ApJ...630..332F} Fox, A.~J., Wakker, B.~P., 
Savage, B.~D., Tripp, T.~M., Sembach, K.~R., \& Bland-Hawthorn, J.\ 2005, 
\apj, 630, 332 

\bibitem[Fox et al.(2006)]{2006ApJS..165..229F} Fox, A.~J., Savage, B.~D., 
\& Wakker, B.~P.\ 2006, \apjs, 165, 229 

\bibitem[Furlanetto et al.(2005)]{2005MNRAS.359..295F} Furlanetto, S.~R., 
Phillips, L.~A., \& Kamionkowski, M.\ 2005, \mnras, 359, 295 

\bibitem[Gaetz \& Salpeter(1983)]{1983ApJS...52..155G} Gaetz, T.~J., \& 
Salpeter, E.~E.\ 1983, \apjs, 52, 155 

\bibitem[Gnat \& Sternberg(2004)]{2004ApJ...608..229G} Gnat, O., \& 
Sternberg, A.\ 2004, \apj, 608, 229 

\bibitem[Heckman et al.(2002)]{2002ApJ...577..691H} Heckman, T.~M., Norman, 
C.~A., Strickland, D.~K., \& Sembach, K.~R.\ 2002, \apj, 577, 691 

\bibitem[Hindmarsh(1983)]{05}Hindmarsh, A.~C., \ 1983, ODEPACK, A Systematized Collection of 
ODE Solvers, in Scientific Computing, R. S. Stepleman et al. (eds.), North-Holland, 
Amsterdam, 1983 (vol. 1 of IMACS Transactions on Scientific Computation), pp. 55-64.

\bibitem[House(1964)]{1964ApJS....8..307H} House, L.~L.\ 1964, \apjs, 8, 
307 

\bibitem[Jordan(1969)]{1969MNRAS.142..501J} Jordan, C.\ 1969, \mnras, 142, 
501 

\bibitem[Jura \& Dalgarno(1972)]{1972ApJ...174..365J} Jura, M., \& 
Dalgarno, A.\ 1972, \apj, 174, 365 

\bibitem[Kafatos(1973)]{1973ApJ...182..433K} Kafatos, M.\ 1973, \apj, 182, 
433 

\bibitem[Kahn(1976)]{1976A&A....50..145K} Kahn, F.~D.\ 1976, \aap, 50, 145 

\bibitem[Kingdon \& Ferland(1996)]{1996ApJS..106..205K} Kingdon, J.~B., \& 
Ferland, G.~J.\ 1996, \apjs, 106, 205 

\bibitem[Landi \& Landini(1999)]{1999A&A...347..401L} Landi, E., \& 
Landini, M.\ 1999, \aap, 347, 401 

\bibitem[Landini \& Fossi(1991)]{1991A&AS...91..183L} Landini, M., \&
Fossi, B.~C.~M.\ 1991, \aaps, 91, 183

\bibitem[Landini \& Monsignori Fossi(1990)]{1990A&AS...82..229L}
Landini, M., \& Monsignori Fossi, B.~C.\ 1990, \aaps, 82, 229

\bibitem[Maller \& Bullock(2004)]{2004MNRAS.355..694M} Maller, A.~H., \& 
Bullock, J.~S.\ 2004, \mnras, 355, 694 

\bibitem[McCray(1987)]{1987sap..book..255M}McCray, R. 1987, 
in Spectroscopy of Astrophysical Plasmas,
Eds.~A.~Dalgarno \& D.~Layzer (Cambridge University Press), p. 255

\bibitem[Mitnik \& Badnell(2004)]{2004A&A...425.1153M} Mitnik, D.~M., \&
Badnell, N.~R.\ 2004, \aap, 425, 1153

\bibitem[Netzer et al.(2005)]{2005ApJ...629..739N} Netzer, H., Lemze, D., 
Kaspi, S., George, I.~M., Turner, T.~J., Lutz, D., Boller, T., \& 
Chelouche, D.\ 2005, \apj, 629, 739 

\bibitem[Nicastro et al.(2005)]{2005Natur.433..495N} Nicastro, F., et al.\ 
2005, \nat, 433, 495 

\bibitem[Ostriker \& Silk(1973)]{1973ApJ...184L.113O} Ostriker, J., \& 
Silk, J.\ 1973, \apjl, 184, L113 

\bibitem[Osterbrock(1989)]{1989agna.book.....O} Osterbrock, D.~E.\ 1989, 
Astrophysics of Gaseous Nebulae \& Active Galactic Nuclei,
~(Mill Valley: University Science Books)

\bibitem[Pequignot et al.(1991)]{1991A&A...251..680P} Pequignot, D.,
Petitjean, P., \& Boisson, C.\ 1991, \aap, 251, 680

\bibitem[Rasmussen et al.(2006)]{2006astro.ph..4515R} Rasmussen, A.~P., 
Kahn, S.~M., Paerels, F., Willem den Herder, J., Kaastra, J., \& de Vries, 
C.\ 2006, astro-ph/0604515 

\bibitem[Raymond et al.(1976)]{1976ApJ...204..290R} Raymond, J.~C., Cox, 
D.~P., \& Smith, B.~W.\ 1976, \apj, 204, 290 

\bibitem[Richter et al.(2004)]{2004ApJS..153..165R} Richter, P., Savage, 
B.~D., Tripp, T.~M., \& Sembach, K.~R.\ 2004, \apjs, 153, 165 

\bibitem[Savage et al.(2002)]{2002ApJ...564..631S} Savage, B.~D., Sembach, 
K.~R., Tripp, T.~M., \& Richter, P.\ 2002, \apj, 564, 631 

\bibitem[Savage et al.(2005)]{2005ApJ...626..776S} Savage, B.~D., Lehner, 
N., Wakker, B.~P., Sembach, K.~R., \& Tripp, T.~M.\ 2005, \apj, 626, 776 

\bibitem[Schmelz et al.(2005)]{2005ApJ...634L.197S} Schmelz, J.~T., 
Nasraoui, K., Roames, J.~K., Lippner, L.~A., \& Garst, J.~W.\ 2005, \apjl, 
634, L197 

\bibitem[Schmutzler \& Tscharnuter(1993)]{1993A&A...273..318S} Schmutzler, 
T., \& Tscharnuter, W.~M.\ 1993, \aap, 273, 318 

\bibitem[Schwarz et al.(1972)]{06}Schwarz, J., McCray, R., \& Stein, R.F.\ 1972, \apj, 175, 673

\bibitem[Sembach \& Savage(1992)]{1992ApJS...83..147S} Sembach, K.~R., \& 
Savage, B.~D.\ 1992, \apjs, 83, 147 

\bibitem[Sembach et al.(2004)]{2004ApJS..155..351S} Sembach, K.~R., Tripp, 
T.~M., Savage, B.~D., \& Richter, P.\ 2004, \apjs, 155, 351 

\bibitem[Shapiro \& Moore(1976)]{1976ApJ...207..460S} Shapiro, P.~R., \& 
Moore, R.~T.\ 1976, \apj, 207, 460 

\bibitem[Shull \& van Steenberg(1982)]{1982ApJS...48...95S} Shull, J.~M., 
\& van Steenberg, M.\ 1982, \apjs, 48, 95 

\bibitem[Shull et al.(2003)]{2003ApJ...594L.107S} Shull, J.~M., Tumlinson, 
J., \& Giroux, M.~L.\ 2003, \apjl, 594, L107 

\bibitem[Slavin et al.(1993)]{1993ApJ...407...83S} Slavin, J.~D., Shull, 
J.~M., \& Begelman, M.~C.\ 1993, \apj, 407, 83 

\bibitem[Smith et al.(1996)]{1996ApJ...473..864S} Smith, R.~K., Krzewina, 
L.~G., Cox, D.~P., Edgar, R.~J., \& Miller, W.~W.~I.\ 1996, \apj, 473, 864 

\bibitem[Soltan et al.(2005)]{2005AA...436...67}Soltan, A.~M., Freyberg, M.~J.,
\& Hasinger, G.\ 2005, \aa, 436, 67

\bibitem[Spitzer(1996)]{1996ApJ...458L..29S} Spitzer, L.~J.\ 1996, \apjl, 
458, L29 

\bibitem[Stancil et al.(1998)]{1998ApJ...502.1006S} Stancil, P.~C., et al.\ 
1998, \apj, 502, 1006 

\bibitem[Sternberg et al.(2002)]{2002ApJS..143..419S} Sternberg, A., McKee, 
C.~F., \& Wolfire, M.~G.\ 2002, \apjs, 143, 419 

\bibitem[Sutherland \& Dopita(1993)]{1993ApJS...88..253S} Sutherland, 
R.~S., \& Dopita, M.~A.\ 1993, \apjs, 88, 253 

\bibitem[Tripp et al.(2000)]{2000ApJ...534L...1T} Tripp, T.~M., Savage, 
B.~D., \& Jenkins, E.~B.\ 2000, \apjl, 534, L1 

\bibitem[Tucker \& Gould(1966)]{1966ApJ...144..244T} Tucker, W.~H., \& 
Gould, R.~J.\ 1966, \apj, 144, 244 

\bibitem[Verner et al.(1996)]{1996ApJ...465..487V} Verner, D.~A.,
Ferland, G.~J., Korista, K.~T., \& Yakovlev, D.~G.\ 1996, \apj, 465, 487

\bibitem[Voronov(1997)]{1997ADNDT..65....1V} Voronov, G.~S.\ 1997,
Atomic Data and Nuclear Data Tables, 65, 1

\bibitem[Williams et al.(2006)]{2006ApJ...642L..95W} Williams, R.~J., 
Mathur, S., Nicastro, F., \& Elvis, M.\ 2006, \apjl, 642, L95 

\bibitem[Zatsarinny et al.(2003)]{2003A&A...412..587Z} Zatsarinny, O.,
Gorczyca, T.~W., Korista, K.~T., Badnell, N.~R., \& Savin, D.~W.\ 2003,
\aap, 412, 587

\bibitem[Zatsarinny et al.(2004)]{2004A&A...417.1173Z} Zatsarinny, O.,
Gorczyca, T.~W., Korista, K.~T., Badnell, N.~R., \& Savin, D.~W.\ 2004a,
\aap, 417, 1173

\bibitem[Zatsarinny et al.(2004)]{2004A&A...426..699Z} Zatsarinny, O.,
Gorczyca, T.~W., Korista, K., Badnell, N.~R., \& Savin, D.~W.\ 2004b,
\aap, 426, 699

\bibitem[Zatsarinny et al.(2005)]{2005A&A...438..743Z} Zatsarinny, O.,
Gorczyca, T.~W., Korista, K.~T., Fu, J., Badnell, N.~R., Mitthumsiri,
W., \& Savin, D.~W.\ 2005a, \aap, 438, 743

\bibitem[Zatsarinny et al.(2005)]{2005A&A...440.1203Z} Zatsarinny, O.,
Gorczyca, T.~W., Korista, K.~T., Fu, J., Badnell, N.~R., Mitthumsiri,
W., \& Savin, D.~W.\ 2005b, \aap, 440, 1203

\bibitem[Zatsarinny et al.(2006)]{2006A&A...447..379Z} Zatsarinny, O.,
Gorczyca, T.~W., Fu, J., Korista, K.~T., Badnell, N.~R., \& Savin, D.~W.\ 2006, \aap, 447, 379

\bibitem[Zsarg{\'o} et al.(2003)]{2003ApJ...586.1019Z} Zsarg{\'o}, J., 
Sembach, K.~R., Howk, J.~C., \& Savage, B.~D.\ 2003, \apj, 586, 1019 
\end{thebibliography}
\end{document}